\documentclass[lettersize,journal]{IEEEtran}
\usepackage{amsmath,amsfonts}
\usepackage{amssymb}
\usepackage{algorithmic}
\usepackage{algorithm}
\usepackage{array}
\usepackage[caption=false,font=normalsize,labelfont=sf,textfont=sf]{subfig}
\usepackage{textcomp}
\usepackage{stfloats}
\usepackage{url}
\usepackage{verbatim}
\usepackage{graphicx}
\usepackage{cite}
\usepackage{hyperref}

\newcounter{MYtempeqncnt}

\newtheorem{proposition}{\bf Proposition}
\newtheorem{theorem}{\bf Theorem}
\newtheorem{remark}{\bf Remark}
\newtheorem{lemma}{\bf Lemma}

\usepackage{bm}
\def\BibTeX{{\rm B\kern-.05em{\sc i\kern-.025em b}\kern-.08em
    T\kern-.1667em\lower.7ex\hbox{E}\kern-.125emX}}

\begin{document}
\title{Block-Level MU-MISO Interference Exploitation Precoding: Optimal Structure and Explicit Duality}

\author{Junwen~Yang,~\IEEEmembership{Graduate~Student~Member,~IEEE,} Ang~Li,~\IEEEmembership{Senior~Member,~IEEE,}	Xuewen~Liao,~\IEEEmembership{Member,~IEEE,}			Christos~Masouros,~\IEEEmembership{Fellow,~IEEE,}~and~A.~L.~Swindlehurst,~\IEEEmembership{Fellow,~IEEE}

\thanks{Manuscript received XXXX; revised XXXX. \emph{(Corresponding authors: Xuewen Liao; Ang Li.)}}

\thanks{J. Yang and A. Li are with the School of Information and Communications Engineering, Faculty of Electronic and Information Engineering, Xi'an Jiaotong University, Xi'an, Shaanxi 710049, China (e-mail: jwyang@stu.xjtu.edu.cn; ang.li.2020@xjtu.edu.cn).}
\thanks{X. Liao is with the School of Information and Communications Engineering, Faculty of Electronic and Information Engineering, Xi’an Jiaotong University, Xi’an, Shaanxi 710049, China, and also with the National Mobile Communications Research Laboratory, Southeast University, Nanjing 210096, China (e-mail: yeplos@mail.xjtu.edu.cn).}
\thanks{C. Masouros is with the Department of Electronic and Electrical Engineering, University College London, London WC1E 7JE, U.K. (e-mail: c.masouros@ucl.ac.uk).}
\thanks{A. L. Swindlehurst is with the Center for Pervasive Communications and Computing, University of California, Irvine, CA 92697, USA (e-mail: swindle@uci.edu).}}

\maketitle

\begin{abstract}
This paper investigates block-level interference exploitation (IE) precoding for multi-user multiple-input single-output (MU-MISO) downlink systems. To overcome the need for symbol-level IE precoding to frequently update the precoding matrix, we propose to jointly optimize all the precoders or transmit signals within a transmission block. The resultant precoders only need to be updated once per block, and while not necessarily constant over all the symbol slots, we refer to the technique as block-level slot-variant IE precoding. Through a careful examination of the optimal structure and the explicit duality inherent in block-level power minimization (PM) and signal-to-interference-plus-noise ratio (SINR) balancing (SB) problems, we discover that the joint optimization can be decomposed into subproblems with smaller variable sizes. As a step further, we propose block-level slot-invariant IE precoding by adding a structural constraint on the slot-variant IE precoding to maintain a constant precoder throughout the block. A novel linear precoder for IE is further presented, and we prove that the proposed slot-variant and slot-invariant IE precoding share an identical solution when the number of symbol slots does not exceed the number of users. Numerical simulations demonstrate that the proposed precoders achieve a significant complexity reduction compared against benchmark schemes, without sacrificing performance. 
\end{abstract}


\section{Introduction}
\IEEEPARstart{W}{ith} the ever-growing demand for high data rates and massive access in wireless communications, independent data streams need to be spatially multiplexed in the same resource block via multiple-input multiple-output (MIMO) or multi-user multiple-input single-output (MU-MISO) systems \cite{6736761,6798744,1197843}. Transmit precoding/beamforming is a predominant interference management technology for MIMO communications \cite{1634819,1310319}. Linear precoding methods leverage only the channel state information (CSI) to design the precoder, assuming the data symbols across different symbol slots are independent and identically distributed (i.i.d.) Gaussian random variables. For example, zero forcing (ZF) \cite{1207369} and regularized ZF (RZF) precoding \cite{1391204} use (regularized) channel inversion to eliminate interference, but the inversion operation results in limited communication performance. There are also nonlinear precoding methods such as dirty-paper precoding (DPC) \cite{1056659}, Tomlinson-Harashima precoding (THP) \cite{1310319}, and vector perturbation (VP) precoding \cite{1413598}, where the precoded signal is a nonlinear combination of the data symbols. The complexity of the above nonlinear precoding methods limit however their practical implementation \cite{1056659,1310319,1413598}.

To improve interference management, optimal linear precoding has been proposed, which models the precoding task as an optimization problem. A variety of optimization techniques have been investigated to seek solutions under different optimality criteria. Power minimization (PM) linear precoding can minimize the transmit power while guaranteeing given quality of service (QoS) conditions, which is commonly expressed in terms of signal-to-interference-plus-noise ratio (SINR) constraints \cite{bengtsson1999optimal,bengtsson2002optimum}. SINR balancing (SB) linear precoding aims to provide fairness among users by maximizing the minimum SINR of all the users in the system, usually subject to a transmit power constraint \cite{1561584,4443878}.

The aforementioned interference management methods treat the interference as a harmful component that should be eliminated. However, recent studies on IE have provided the novel insight that the multiuser interference is beneficial when it is constructive, and can be exploited to improve detection performance. This technique is also referred to as constructive interference (CI) precoding \cite{4286950,4801492,5605266,7042789,7103338,9035662}. IE precoding is data symbol dependent, and the precoder is commonly updated on a symbol-by-symbol basis, so it is often referred to as symbol-level precoding (SLP) \cite{7042789,9035662,7404291,7942010,8477154,8575164,8170315,8359237}. Recently, the idea of IE via SLP has been studied in various emerging wireless communication fields, such as reconfigurable intelligent surface (RIS) \cite{9435988,9219206}, low-resolution digital-to-analog converters (DACs) \cite{9007045,9446685}, and integrated sensing and communications (ISAC) \cite{8355705,9534484}.

As promising as the symbol-level IE precoding is, its need to update the precoder in each symbol slot imposes implementation challenges that originate from the SLP design criteria. Based on this observation, a block-level transmit power constrained SB IE precoding algorithm was proposed in \cite{9962829}, where a constant IE precoder is designed for an entire transmission block/frame.

Given that the precoded symbol-level IE signal can be viewed as a nonlinear combination of data symbols on the block level, SLP is commonly classified as nonlinear precoding \cite{9712356}. The CI-BLP proposed in \cite{9962829} offers a potential linear method for IE precoding, since it designs a unified constant precoder for a collection of data symbols over a transmission block. However, the work of \cite{9962829} primarily concentrates on the SB problem and specifically addresses the scenario where the number of symbol slots exceeds the number of users. It thus does not provide proofs for the generic existence of linear precoding for IE.

To further exploit the potential of block-level optimizations for IE, in this paper we propose block-level slot-variant and slot-invariant IE precoding for both PM and SB problems. Here, "block-level" refers to the fact that the precoding optimization is solved once per block, while "slot-variant" indicates that the precoding solution can be considered as a time-varying linear transformation of the data symbols. The main contributions of this paper are summarized below:
\begin{itemize}
\item \emph{Block-Level Slot-Variant IE Precoding:} We propose block-level PM and weighted SB slot-variant IE precoders. In a single transmission block/frame, multiple precoders or equivalently transmit signals are optimized with block-level performance metrics. The optimal structure of the precoders is derived for both PM and (weighted) SB problems. We further prove that the block-level PM slot-variant IE precoding and the PM-SLP share the same optimal solution, and the relationship between SB slot-variant IE precoding and SB-SLP is also derived.

\item \emph{Block-Level Slot-Invariant IE Precoding:} By introducing a structural constraint on the slot-invariant IE precoding matrix, we obtain PM and weighted SB IE precoders that are invariant across the symbol slots within a transmission block/frame, and hence become block-level slot-invariant IE precoders. Their optimal structure is revealed by deriving the Lagrangian dual problems, while a novel explicit duality is revealed in the proof of the solutions to the two slot-invariant IE precoding problems.

\item \emph{Existence of Linear Precoding for IE:} On the premise that the number of symbol slots does not exceed the number of users, we prove the existence of a constant precoder over different symbol slots for arbitrary precoding, and reveal that the IE gain can be achieved by a linear precoder. This is contrary to the common assumption that IE precoders must be calculated in each symbol slot to fully exploit the multiuser interference.
\end{itemize}
Extensive numerical simulations are conducted to demonstrate the superiority of the proposed block-level slot-variant and slot-invariant IE precoding. The simulations also validate our derivations of the optimality of the linear precoder.

The remainder of this paper is organized as follows. Section \ref{secModel} introduces the system model and preliminaries. Section \ref{secSVP} presents the block-level PM/SB slot-variant IE precoding. Section \ref{secSIP} investigates the block-level PM and weighted SB slot-invariant IE precoding. Section \ref{secLinear} proposes the novel linear precoder for IE. Section \ref{secResults} presents the numerical results, and Section \ref{secConclusion} concludes the paper.

\textit{Notation:} Scalars, vectors, and matrices are denoted by plain lower-case, bold lower-case, and bold capital letters, respectively. $(\cdot)^T$, $(\cdot)^H$, and $(\cdot)^{-1}$ denote transpose, conjugate transpose, and inverse operators, respectively. $\mathbb{C}^{M\times N}$ and $\mathbb{R}^{M\times N}$ denote the sets of $M\times N$ matrices with complex and real entries, respectively. $\left| \cdot \right|$ represents the absolute value of a real scalar or the modulus of a complex scalar. $\left\|\cdot\right\|$ denotes the Euclidean norm of a vector or spectral norm of a matrix. $\Re \{\cdot\}$ and $\Im \{\cdot\}$ respectively denote the real part and imaginary part of a complex input. $\succeq$ denotes element-wise inequality. $\mathbf{0}$, $\mathbf{1}$, and $\mathbf{I}$ represent respectively, the all-zeros vector, the all-ones vector, and the identity matrix. $\oslash$ denotes the element-wise division. $\text{diag}\{\cdot\}$ returns a diagonal matrix with the entries of the input vector on the main diagonal.

\section{System Model and Preliminaries}
\label{secModel}
\subsection{System Model}
We consider a time-division duplex (TDD) MU-MISO downlink system with $N_t$ transmit antennas and $N_r$ single-antenna users. Assume the channel is constant over a transmission block/frame, which consists of $N_s$ symbol slots. The received signal $\tilde{y}_{k,i}$ at the $k$-th user in the $i$-th symbol slot can be represented as
\begin{align}
\label{eqModel}
\tilde{y}_{k,i}&=\underbrace{\tilde{\mathbf{h}}^T_{k}\tilde{\mathbf{w}}^k_i\tilde{s}_{k,i}}_{\text{desired signal}}+\underbrace{\sum_{g=1,g\neq k}^{N_r}\tilde{\mathbf{h}}^T_{k}\tilde{\mathbf{w}}^g_i\tilde{s}_{g,i}}_{\text{interference}}+\underbrace{\tilde{z}_{k,i}}_{\text{noise}} \nonumber\\
&=\tilde{\mathbf{h}}^T_{k}\tilde{\mathbf{W}}_i\tilde{\mathbf{s}}_{:,i}+\tilde{z}_{k,i} =\tilde{\mathbf{h}}^T_{k}\tilde{\mathbf{x}}_i+\tilde{z}_{k,i} ,
\end{align}
where $\tilde{\mathbf{h}}_{k}\in\mathbb{C}^{N_t}$ represents the channel between the transmitter and the $k$-th user; $\tilde{s}_{k,i}$ and $\tilde{z}_{k,i}\sim \mathcal{CN}(0, \sigma^2)$ denote the modulated data symbol and the additive white Gaussian noise of the $k$-th user in the $i$-th symbol slot; $\tilde{\mathbf{W}}_i\triangleq\begin{bmatrix}
\tilde{\mathbf{w}}^1_i & \cdots & \tilde{\mathbf{w}}^{N_{r}}_i
\end{bmatrix}\in\mathbb{C}^{N_t\times N_r}$, $\tilde{\mathbf{s}}_{:,i}\triangleq\left[\tilde{s}_{k,i},\cdots,\tilde{s}_{N_r,i}\right]^T$, and $\tilde{\mathbf{x}}_i\triangleq \tilde{\mathbf{W}}_i\tilde{\mathbf{s}}_{:,i}$ denote the precoder, the data symbol vector, and the precoded transmit signal in the $i$-th symbol slot, respectively. We assume phase-shift keying (PSK) signaling in this paper, but the IE design can be readily extended to quadrature amplitude modulation (QAM) \cite{10229497,9962829}. The term `precoding technique' or `precoder design' refers to the computation of the precoder $\tilde{\mathbf{W}}_i$, and it can also mean the direct computation of the transmit signal vector $\tilde{\mathbf{x}}_i$. Given the transmit signal vector $\tilde{\mathbf{x}}_i$ and any data symbol vector $\tilde{\mathbf{s}}_{:,i}$, the precoder $\tilde{\mathbf{W}}_i$ can be obtained by the following expression \cite{10229497}:
\begin{equation}
\label{eqWx}
\tilde{\mathbf{W}}_i=\tilde{\mathbf{x}}_i\frac{\tilde{\mathbf{s}}^H_{:,i}}{\tilde{\mathbf{s}}^H_{:,i}\tilde{\mathbf{s}}_{:,i}}.
\end{equation}

The precoding matrix varies with the symbol slots when slot-variant BLP or SLP is adopted. When slot-invariant BLP is employed at the transmitter, a constant precoding matrix will be used to precode all the $N_r\times N_s$ data symbols transmitted in one transmission block/frame. In such cases, we omit the subscript of the block-level slot-invariant precoder, i.e., $\tilde{\mathbf{W}}_i=\tilde{\mathbf{W}}_{j\neq i}=\tilde{\mathbf{W}}$. 

\subsection{Interference Exploitation}
\begin{figure}[!t]
\centering
\includegraphics[]{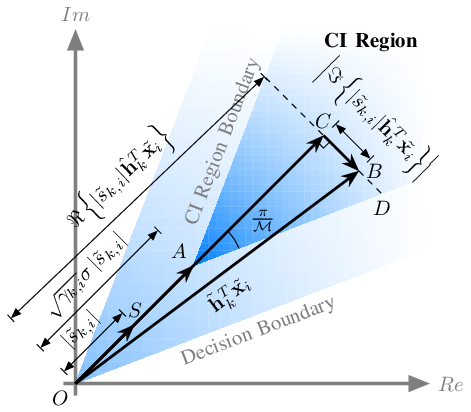}
\caption{Geometric interpretation of IE for a generic $\mathcal{M}$-PSK constellation.}
\label{fig_CI}
\end{figure}

The interference term $\sum_{g=1,g\neq k}^{N_r}\tilde{\mathbf{h}}^T_{k}\tilde{\mathbf{W}}_i\tilde{s}_{g,i}$ in \eqref{eqModel} is often treated as noise to be eliminated or suppressed. On the contrary, with the aid of CSI and knowledge of the data symbols, IE precoding manages to convert the interference part of the received signal into CI by optimizing the noiseless received signal \cite{7103338}. To measure the quality of service (QoS) of a specific IE precoding technique, we introduce the instantaneous SINR of the $k$-th user in the $i$-th symbol slot as
$
\mathrm{SINR}_{k,i}\triangleq\frac{\left|\tilde{\mathbf{h}}^T_{k}\tilde{\mathbf{x}}_i\right|^2}{\sigma^2}
$
\cite{7103338}, which can also be viewed as an indirect indicator of the degree of IE.

It should be noted that the instantaneous SINR is effective only when the interference is exploited and the noiseless received signal is pushed deeper into the detection region. Fig. \ref{fig_CI} gives a geometric interpretation of the idea of IE for a generic $\mathcal{M}$-PSK constellation, where $\overrightarrow{OS}=\tilde{s}_{k,i}$, $\overrightarrow{OA}=\sqrt{\gamma_{k,i}}\sigma\tilde{s}_{k,i}$, and $\overrightarrow{OB}=\tilde{\mathbf{h}}^T_{k}\tilde{\mathbf{x}}_i$ denote the data symbol, the nominal received data symbol with instantaneous SINR threshold $\mathrm{SINR}_{k,i}\geq\gamma_{k,i}\left|\tilde{s}_{k,i}\right|^2$, and the noiseless received signal, respectively. We can observe that as long as the noiseless received signal $\overrightarrow{OB}$ lies in the CI region, IE is achieved with a certain SINR threshold. Leveraging the method proposed in \cite{10229497}, we orthogonally decompose $\overrightarrow{OB}$ along the direction of $\overrightarrow{OS}$, then we have $\overrightarrow{OB}=\overrightarrow{OC}+\overrightarrow{CB}$, where $\overrightarrow{CB}$ intersects with the nearest CI region boundary at point $D$. In this way, we can readily formulate the condition for the received signal to be located in the CI region as $\left|\overrightarrow{CD}\right|\geq\left|\overrightarrow{CB}\right|$. After some transformations, the CI constraints for the IE system can be written in terms of the instantaneous SINR threshold as \cite{7103338}
\begin{equation}
\Re\left\{\hat{\mathbf{h}}^T_{k}\tilde{\mathbf{x}}_i\right\}-\frac{\left|\Im\left\{\hat{\mathbf{h}}^T_{k}\tilde{\mathbf{x}}_i\right\}\right|}{\tan \frac{\pi}{\mathcal{M}}}\geq\sqrt{\gamma_{k,i}}\sigma,\, \forall k,
\end{equation}
where $\hat{\mathbf{h}}^T_{k}\triangleq\frac{\tilde{\mathbf{h}}^T_{k}}{\tilde{s}_{k,i}}$. Since the instantaneous SINR thresholds are incorporated into the CI constraints, we use the terms SINR constraints and CI constraints interchangeably in this paper.

\section{Block-Level Slot-Variant Interference-Exploitation Precoder}
\label{secSVP}
In this section, we propose a slot-variant IE precoder that only needs to compute the precoders or transmit signal vectors once per transmission block/frame while coherently considering the communication performance for all the symbol slots. Moreover, the proposed block-level optimizations can achieve long-term operation, instead of focusing only on the short-term symbol-level design metrics.
\subsection{PM Slot-Variant IE Precoder}
To design a slot-variant IE precoder for the PM problem, the transmitter aims to minimize the block-level transmit power while satisfying the prescribed SINR constraints for the noiseless received signal of all the $N_r$ users over each symbol slot. Therefore, the symbol-level transmit power objective function in the PM-SLP problem mentioned in \cite{7103338} is relaxed to block-level transmit power. This allows for a joint optimization of the precoder over a collection of symbol slots within a transmission block. It can be formulated in each transmission block as
\begin{align}
\label{eqBLPMx}
\min_{\left\{\tilde{\mathbf{x}}_i\right\}} &~ \sum_{i=1}^{N_s}\left\|\tilde{\mathbf{x}}_i\right\|^2 \nonumber\\
\text{s.t.} &~ \Re\left\{\hat{\mathbf{h}}^T_{k}\tilde{\mathbf{x}}_i\right\}-\frac{\left|\Im\left\{\hat{\mathbf{h}}^T_{k}\tilde{\mathbf{x}}_i\right\}\right|}{\tan \frac{\pi}{\mathcal{M}}}\geq\sqrt{\gamma_{k,i}}\sigma,\, \forall k,\forall i.
\end{align}
The formulation in \eqref{eqBLPMx} is a linearly constrained quadratic programming problem, and thus is convex. Although it can be solved using standard optimization tools like CVX, efficiently addressing it with customized algorithms poses challenges. The first and foremost obstacle is the complex-valued optimization variables and sophisticated constraint structure. Therefore we reformulate problem \eqref{eqBLPMx} into a real-valued equivalent form with explicit linear constraints:
\begin{align}
\label{eqBLPMxRe}
\min_{\left\{\mathbf{x}_i\right\}} ~ \sum_{i=1}^{N_s}\left\|\mathbf{x}_i\right\|^2 ~
\text{s.t.} ~ \mathbf{T}\hat{\mathbf{S}}_i\mathbf{H}\mathbf{x}_i\succeq\mathbf{b}_i,\, \forall i,
\end{align}
where 
\begin{subequations}
\label{eqMtxDef}
\begin{align}
\mathbf{x}_i\triangleq&\begin{bmatrix}
\Re \left\{\tilde{\mathbf{x}}_i\right\} \\
\Im \left\{\tilde{\mathbf{x}}_i\right\} 
\end{bmatrix}\in\mathbb{R}^{2N_t},\forall i, \\
\mathbf{T}\triangleq &\begin{bmatrix}
\mathbf{I}&-\frac{1}{\tan\frac{\pi}{\mathcal{M}}}\mathbf{I}\\
\frac{1}{\tan\frac{\pi}{\mathcal{M}}}\mathbf{I}&\mathbf{I}
\end{bmatrix}\in\mathbb{R}^{2N_r\times 2N_r},\\
\hat{\mathbf{S}}_i\triangleq &\begin{bmatrix}
\Re\left\{\tilde{\mathbf{S}}_i\right\}&-\Im\left\{\tilde{\mathbf{S}}_i\right\}\\
\Im\left\{\tilde{\mathbf{S}}_i\right\}&\Re\left\{\tilde{\mathbf{S}}_i\right\}
\end{bmatrix}\in\mathbb{R}^{2N_r \times 2N_r},\forall i,\\
\tilde{\mathbf{S}}_i \triangleq & \text{diag}\left\{1\oslash\tilde{\mathbf{s}}_{:,i}\right\}\in\mathbb{R}^{N_r \times N_r},\forall i,\\
\mathbf{H} \triangleq &\begin{bmatrix}
\Re\left\{\tilde{\mathbf{H}}\right\}&-\Im\left\{\tilde{\mathbf{H}}\right\}\\
\Im\left\{\tilde{\mathbf{H}}\right\}&\Re\left\{\tilde{\mathbf{H}}\right\}
\end{bmatrix}\in\mathbb{R}^{2N_r \times 2N_t},\\
\tilde{\mathbf{H}} \triangleq &\begin{bmatrix}
\tilde{\mathbf{h}}_{1} & \cdots & \tilde{\mathbf{h}}_{N_s}
\end{bmatrix}^{T}\in\mathbb{R}^{N_r \times N_t},\\
\mathbf{b}_i \triangleq &\begin{bmatrix}
\tilde{\mathbf{b}}^T_i & \tilde{\mathbf{b}}^T_i
\end{bmatrix}^{T}\in \mathbb{R}^{2N_r}, \forall i,\\
\tilde{\mathbf{b}}_i \triangleq &\begin{bmatrix}
\sqrt{\gamma_{1,i}}\sigma & \cdots & \sqrt{\gamma_{N_r,i}}\sigma
\end{bmatrix}^{T}\in \mathbb{R}^{N_r}, \forall i.
\end{align}
\end{subequations}
The real-valued PM problem for the slot-variant IE precoder has a simplified structure compared with the complex-valued one. It is however still not trivial to obtain the optimal solution due to the irregular polyhedral feasible region in \eqref{eqBLPMxRe}. Since its convexity guarantees that the duality gap is zero, we can derive the optimal solution structure of \eqref{eqBLPMxRe} by further formulating its dual problem and leveraging Lagrangian duality.

\subsection{Optimal Structure for PM Slot-Variant IE Precoder}
To derive the optimal precoder structure, we first write the Lagrangian function of \eqref{eqBLPMxRe}:
\begin{align}
\mathcal{L}\left(\left\{\mathbf{x}_i\right\}, \left\{\boldsymbol{\lambda}_i\right\}\right)\triangleq \sum_{i=1}^{N_s}\left\|\mathbf{x}_i\right\|^2+\boldsymbol{\lambda}^T_i\left(\mathbf{b}_i-\mathbf{T}\hat{\mathbf{S}}_i\mathbf{H}\mathbf{x}_i\right),
\end{align}
where $\boldsymbol{\lambda}_i$ is the Lagrangian dual variable associated with the CI constraints in the $i$-th symbol slot. The Karush-Kuhn-Tucker (KKT) optimal conditions are given by
\begin{align}
\mathbf{T}\hat{\mathbf{S}}_i\mathbf{H}\mathbf{x}_i \succeq &\mathbf{b}_i,\, \forall i,\\
\boldsymbol{\lambda}_i \succeq &\mathbf{0}, \forall i,\\
\boldsymbol{\lambda}_{i,k}\left(\mathbf{b}_i-\mathbf{T}\hat{\mathbf{S}}_i\mathbf{H}\mathbf{x}_i\right)_{k} =& 0, \forall k, \forall i,\\
\label{eqStationSVPM}
\frac{\partial \mathcal{L}\left(\left\{\mathbf{x}_i\right\}, \left\{\boldsymbol{\lambda}_i\right\}\right)}{\partial \mathbf{x}_i} =& \mathbf{0}, \forall i,
\end{align}
where $\boldsymbol{\lambda}_{i,k}$ and $\left(\mathbf{b}_i-\mathbf{T}\hat{\mathbf{S}}_i\mathbf{H}\mathbf{x}_i\right)_{k}$ denote the $k$-th entry of $\boldsymbol{\lambda}_{i}$ and $\mathbf{b}_i-\mathbf{T}\hat{\mathbf{S}}_i\mathbf{H}\mathbf{x}_i$, respectively. From \eqref{eqStationSVPM} we have
\begin{align}
2\mathbf{x}_i-\mathbf{H}^T\mathbf{T}^T\hat{\mathbf{S}}^T_i\boldsymbol{\lambda}_i=\mathbf{0}, \forall i.
\end{align}
Therefore, the optimal solution of \eqref{eqBLPMxRe} is given by
\begin{align}
\label{eqStrucSVPM}
\mathbf{x}_i=\frac{1}{2}\mathbf{H}^T\mathbf{T}^T\hat{\mathbf{S}}^T_i\boldsymbol{\lambda}_i, \forall i.
\end{align}

The dual problem of \eqref{eqBLPMxRe} is to maximize the dual function, i.e., $g\left(\left\{\boldsymbol{\lambda}_i\right\}\right) \triangleq \min_{\left\{\mathbf{x}_i\right\}}\, \mathcal{L}\left(\left\{\mathbf{x}_i\right\}, \left\{\boldsymbol{\lambda}_i\right\}\right)$, subjected to non-negative constraints. It can be written as:
\begin{align}
\label{eqBLPMdualOrig}
\max_{\left\{\boldsymbol{\lambda}_i\right\}} \min_{\left\{\mathbf{x}_i\right\}} ~ \mathcal{L}\left(\left\{\mathbf{x}_i\right\}, \left\{\boldsymbol{\lambda}_i\right\}\right) ~
\text{s.t.} ~ \boldsymbol{\lambda}_i \succeq \mathbf{0}, \forall i.
\end{align}
By substituting the optimal solution structure \eqref{eqStrucSVPM} into the above dual problem, we can reformulate it in the following form:
\begin{align}
\label{eqBLPMdual}
\min_{\left\{\boldsymbol{\lambda}_i\right\}} &~ \sum_{i=1}^{N_s}\frac{1}{4}\boldsymbol{\lambda}^T_i\hat{\mathbf{S}}_i\mathbf{T}\mathbf{H}\mathbf{H}^T\mathbf{T}^T\hat{\mathbf{S}}^T_i\boldsymbol{\lambda}_i-\boldsymbol{\lambda}^T_i\mathbf{b}_i \nonumber\\
\text{s.t.} &~ \boldsymbol{\lambda}_i \succeq \mathbf{0}, \forall i.
\end{align}
This problem is much simpler than the original problem in \eqref{eqBLPMxRe}. Since the polyhedral constraints are transferred to simple nonnegative constraints.

\subsection{SB Slot-Variant IE Precoder}
In designing a block-level slot-variant IE precoder for the SB problem, we aim to attain fairness among all the users over a transmission block by balancing the instantaneous received SINR. It can be seen from Fig. \ref{fig_CI} that the aforementioned rationale can be interpreted as maximizing the minimum amplitude of $\overrightarrow{OA}$ over $N_s$ symbol slots, which is formulated as an optimization problem on the precoding matrix in \eqref{eqSVSBorig} at the top of the next page.

\begin{figure*}[!t]
\normalsize
\setcounter{MYtempeqncnt}{\value{equation}}
\setcounter{equation}{15}
\begin{align}
\label{eqSVSBorig}
\max_{\left\{\tilde{\mathbf{x}}_i\right\}}\min_{k,i}\arg \min &~ \left\{\frac{1}{\sqrt{\gamma_{k,i}}\sigma}\left(\Re\left\{\hat{\mathbf{h}}^T_{k}\tilde{\mathbf{x}}_i\right\}-\frac{\Im\left\{\hat{\mathbf{h}}^T_{k}\tilde{\mathbf{x}}_i\right\}}{\tan \frac{\pi}{\mathcal{M}}}\right),\frac{1}{\sqrt{\gamma_{k,i}}\sigma}\left(\Re\left\{\hat{\mathbf{h}}^T_{k}\tilde{\mathbf{x}}_i\right\}+\frac{\Im\left\{\hat{\mathbf{h}}^T_{k}\tilde{\mathbf{x}}_i\right\}}{\tan \frac{\pi}{\mathcal{M}}}\right)\right\} \nonumber\\
\text{s.t.} &~ \sum_{i=1}^{N_s}\left\|\tilde{\mathbf{x}}_i\right\|^2\leq \sum_{i=1}^{N_s}p_i
\end{align}
\setcounter{equation}{\value{MYtempeqncnt}}
\hrulefill
\vspace*{4pt}
\end{figure*}
\addtocounter{equation}{1}

Since handling such a max-min problem is difficult, we recast it as a maximization problem by introducing an auxiliary variable $t$. The new problem can be formulated as:
\begin{align}
\label{eqBLSBcomplex}
\max_{\left\{\tilde{\mathbf{x}}_i\right\},\,t} &~ t \nonumber\\
\text{s.t.} &~ \Re\left\{\hat{\mathbf{h}}^T_{k}\tilde{\mathbf{x}}_i\right\}-\frac{\left|\Im\left\{\hat{\mathbf{h}}^T_{k}\tilde{\mathbf{x}}_i\right\}\right|}{\tan \frac{\pi}{\mathcal{M}}}\geq t \sqrt{\gamma_{k,i}}\sigma,\, \forall k, \forall i,\nonumber\\
&~ \sum_{i=1}^{N_s}\left\|\tilde{\mathbf{x}}_i\right\|^2\leq \sum_{i=1}^{N_s}p_i,
\end{align}
where $p_i$ denotes the transmit power budget for the $i$-th symbol slot, and $\sum_{i=1}^{N_s}p_i$ denotes the block-level transmit power budget. With a slight abuse of notation, we will let $\frac{1}{\sqrt{\gamma_{k,i}}}$ denote the square root of the weight applied to $\mathrm{SINR}_{k,i}$ in the context of the weighted SB problem. Similar to our manipulations on the PM problem, we again reformulate the above problem into a real-valued form:
\begin{align}
\label{eqBLSBRe}
\min_{\left\{\mathbf{x}_i\right\}, t} &~ -t \nonumber\\
\text{s.t.} &~ \mathbf{T}\hat{\mathbf{S}}_i\mathbf{H}\mathbf{x}_i\succeq t\mathbf{b}_i,\, \forall i, ~
~ \sum_{i=1}^{N_s}\left\|\mathbf{x}_i\right\|^2\leq \sum_{i=1}^{N_s}p_i,
\end{align}
where the definitions follow those in \eqref{eqMtxDef}. This is a quadratically constrained linear programming problem, and thus is convex.
\begin{proposition}
\label{PropActive}
\it The block level transmit power constraint in \eqref{eqBLSBRe} is active when optimality is achieved, i.e.,
\begin{equation}
\label{eqActPow}
\sum_{i=1}^{N_s}\left\|\mathbf{x}_i\right\|^2 = \sum_{i=1}^{N_s}p_i.
\end{equation} 
\end{proposition}
\begin{IEEEproof}
Assume the above proposition does not hold, then the optimal solution satisfies $\sum_{i=1}^{N_s}\left\|\mathbf{x}_i\right\|^2\leq \sum_{i=1}^{N_s}p_i$. We can always find a feasible solution $\left\{\dot{\mathbf{x}}_i\right\}=\alpha \left\{\mathbf{x}_i\right\}, \alpha\geq 1$, such that $\mathbf{T}\hat{\mathbf{S}}_i\mathbf{H}\dot{\mathbf{x}}_i\succeq \alpha t\mathbf{b}_i, \forall i$, which contradicts the optimality of $t$.
\end{IEEEproof}

In the next subsection, we will derive the optimal solution structure and formulate the Lagrangian dual problem.

\subsection{Optimal Structure for SB Slot-Variant IE Precoder}
\label{secStrucSBSV}
The Lagrangian function of \eqref{eqBLSBRe} is defined by
\begin{align}
\label{eqBLSBReLag}
\mathcal{L}\left(\left\{\mathbf{x}_i\right\}, \left\{\boldsymbol{\lambda}_i\right\}, \mu \right) \triangleq & -t+\sum_{i=1}^{N_s}\boldsymbol{\lambda}^T_i\left(t\mathbf{b}_i-\mathbf{T}\hat{\mathbf{S}}_i\mathbf{H}\mathbf{x}_i\right)\nonumber\\
&+\mu\left(\sum_{i=1}^{N_s}\left\|\mathbf{x}_i\right\|^2-\sum_{i=1}^{N_s}p_i\right),
\end{align}
where $\boldsymbol{\lambda}_i$ is the Lagrangian dual variable associated with the CI constraints in the $i$-th symbol slot, and $\mu$ is the Lagrangian dual variable associated with the block-level transmit power constraint.

To derive the Lagrangian dual problem, we write the KKT optimal conditions below:
\begin{align}
\mathbf{T}\hat{\mathbf{S}}_i\mathbf{H}\mathbf{x}_i \succeq & t\mathbf{b}_i,\, \forall i, \\
\sum_{i=1}^{N_s}\left\|\mathbf{x}_i\right\|^2 \leq & \sum_{i=1}^{N_s}p_i,\\
\boldsymbol{\lambda}_i \succeq &\mathbf{0}, \forall i,\\
\label{eqDualFeasSVSB}
\mu \geq & 0,\\
\boldsymbol{\lambda}_{i,k}\left(t\mathbf{b}_i-\mathbf{T}\hat{\mathbf{S}}_i\mathbf{H}\mathbf{x}_i\right)_{k} =& 0, \forall k, \forall i,\\
\label{eqComplSVSB}
\mu\left(\sum_{i=1}^{N_s}\left\|\mathbf{x}_i\right\|^2-\sum_{i=1}^{N_s}p_i\right) = & 0,\\
\label{eqStationSVSBx}
\frac{\partial \mathcal{L}\left(\left\{\mathbf{x}_i\right\}, \left\{\boldsymbol{\lambda}_i\right\}, \mu \right)}{\partial \mathbf{x}_i} =& \mathbf{0}, \forall i,\\
\label{eqStationSVSBt}
\frac{\partial \mathcal{L}\left(\left\{\mathbf{x}_i\right\}, \left\{\boldsymbol{\lambda}_i\right\}, \mu \right)}{\partial t} =& 0,
\end{align}
where $\boldsymbol{\lambda}_{i,k}$ and $\left(t\mathbf{b}_i-\mathbf{T}\hat{\mathbf{S}}_i\mathbf{H}\mathbf{x}_i\right)_{k}$ denote the $k$-th entry of $\boldsymbol{\lambda}_{i}$ and $t\mathbf{1}-\mathbf{T}\hat{\mathbf{S}}_i\mathbf{H}\mathbf{x}_i$, respectively. From Proposition \ref{PropActive}, we see that the block-level transmit power constraint cannot be ignored, therefore $\mu \neq 0$. Then based on the stationarity condition in \eqref{eqStationSVSBx}, we can directly write the optimal solution structure for the block-level SB slot-variant IE precoder as:
\begin{align}
\label{eqStrucSVSB}
\mathbf{x}_i=\frac{1}{2\mu}\mathbf{H}^T\hat{\mathbf{S}}^T_i\mathbf{T}^T\boldsymbol{\lambda}_i, \forall i.
\end{align}
We can subsequently write the dual function as:
\begin{align}
\label{eqDualFuncSVSBorig}
g\left(\left\{\boldsymbol{\lambda}_i\right\}, \mu \right) \triangleq & \min_{\left\{\mathbf{x}_i\right\}}\, \mathcal{L}\left(\left\{\mathbf{x}_i\right\}, t, \left\{\boldsymbol{\lambda}_i\right\}, \mu\right)\nonumber\\
 = &-\frac{1}{4\mu}\sum_{i=1}^{N_s}\boldsymbol{\lambda}^T_i\mathbf{T}\hat{\mathbf{S}}_i\mathbf{H}\mathbf{H}^T\hat{\mathbf{S}}^T_i\mathbf{T}^T\boldsymbol{\lambda}_i-\mu \sum_{i=1}^{N_s}p_i.
\end{align}

Substituting the optimal solution structure \eqref{eqStrucSVSB} into \eqref{eqActPow}, we can write $\mu$ in terms of $\left\{\boldsymbol{\lambda}_{i}\right\}$ as:
\begin{align}
\label{eqMuSVSB}
\mu=\sqrt{\frac{\sum_{i=1}^{N_s}\boldsymbol{\lambda}^T_i\mathbf{T}\hat{\mathbf{S}}_i\mathbf{H}\mathbf{H}^T\hat{\mathbf{S}}^T_i\mathbf{T}^T\boldsymbol{\lambda}_i}{4\sum_{i=1}^{N_s}p_i}}.
\end{align}
Therefore, the optimal solution structure for the block-level SB slot-variant IE precoder in \eqref{eqStrucSVSB} turns out to be:
\begin{align}
\mathbf{x}_i=\sqrt{\frac{\sum_{i=1}^{N_s}p_i}{\sum_{i=1}^{N_s}\boldsymbol{\lambda}^T_i\mathbf{T}\hat{\mathbf{S}}_i\mathbf{H}\mathbf{H}^T\hat{\mathbf{S}}^T_i\mathbf{T}^T\boldsymbol{\lambda}_i}}\mathbf{H}^T\hat{\mathbf{S}}^T_i\mathbf{T}^T\boldsymbol{\lambda}_i, \forall i.
\end{align}

By substituting \eqref{eqMuSVSB} into \eqref{eqDualFuncSVSBorig}, the dual variable $\mu$ therein can be eliminated. Thus the dual function can be reformulated as:
\begin{align}
\label{eqDualFuncSVSB}
g\left(\left\{\boldsymbol{\lambda}_i\right\} \right) \triangleq & \min_{\left\{\mathbf{x}_i\right\}}\, \mathcal{L}\left(\left\{\mathbf{x}_i\right\}, t, \left\{\boldsymbol{\lambda}_i\right\}, \mu\right)\nonumber\\
 = &-\sqrt{\sum_{j=1}^{N_s}p_j\sum_{i=1}^{N_s}\boldsymbol{\lambda}^T_i\mathbf{T}\hat{\mathbf{S}}_i\mathbf{H}\mathbf{H}^T\hat{\mathbf{S}}^T_i\mathbf{T}^T\boldsymbol{\lambda}_i}.
\end{align}
Since the square root function is monotonic, the dual problem of \eqref{eqBLSBRe} can be expressed as:
\begin{align}
\min_{\left\{\boldsymbol{\lambda}_i\right\}} &~  \sum_{i=1}^{N_s}\boldsymbol{\lambda}^T_i\mathbf{T}\hat{\mathbf{S}}_i\mathbf{H}\mathbf{H}^T\hat{\mathbf{S}}^T_i\mathbf{T}^T\boldsymbol{\lambda}_i \nonumber\\
\text{s.t.} &~ \boldsymbol{\lambda}_i \succeq \mathbf{0}, \forall i, ~
~ \sum_{i=1}^{N_s}\mathbf{b}^T_i \boldsymbol{\lambda}_i-1=0.
\end{align}

\subsection{Relationship to Traditional PM/SB-SLP}
\label{secRela}
In this subsection, we begin by proving that the block-level PM slot-variant IE precoding and PM-SLP share an identical optimal solution, then present the explicit duality between block-level PM and SB slot-variant IE precoders, based on which we investigate the connections between block-level SB slot-variant IE precoder and the conventional SB-SLP in \cite{7103338}.

In the $i$-th symbol slot, the PM-SLP approach designs the transmit signal $\tilde{\mathbf{x}}_i$ using \cite{7103338}:
\begin{align}
\label{eqSLPMx}
\min_{\tilde{\mathbf{x}}_i} &~ \left\|\tilde{\mathbf{x}}_i\right\|^2 \nonumber\\
\text{s.t.} &~ \Re\left\{\hat{\mathbf{h}}^T_{k}\tilde{\mathbf{x}}_i\right\}-\frac{\left|\Im\left\{\hat{\mathbf{h}}^T_{k}\tilde{\mathbf{x}}_i\right\}\right|}{\tan \frac{\pi}{\mathcal{M}}}\geq\sqrt{\gamma_{k,i}}\sigma,\, \forall k.
\end{align}
We summarize the connections between the proposed block-level PM slot-variant IE precoder and the existing PM-SLP in the following theorem.

\begin{theorem}
\label{theorEquaSolu}
\it The block-level PM slot-variant IE precoder \eqref{eqBLPMx} and the PM-SLP precoder \eqref{eqSLPMx} have identical optimal solutions and precoding matrices in each symbol slot.
\end{theorem}
\begin{IEEEproof}[\bf Proof]
Note that in \eqref{eqBLPMx}, the constraints on $\tilde{\mathbf{x}}_{i}$ are independent of the choice of other $\tilde{\mathbf{x}}_{j}, \forall j\neq i$, and the contributions of each $\tilde{\mathbf{x}}_{i}$ to the criterion function do not dependent on the other $\tilde{\mathbf{x}}_{j}, \forall j\neq i$. Therefore, we can decompose the block-level PM slot-variant IE precoding problem \eqref{eqBLPMx} into $N_s$ subproblems over the symbol slots and solve them independently, which means that the optimal solutions to the problem for block-level PM slot-variant IE precoder and PM-SLP problem are identical.
\end{IEEEproof}

The weighted SB-SLP problem in the $i$-th symbol slot can be formulated as \cite{10229497,7103338}
\begin{align}
\label{eqSLSB}
\max_{\tilde{\mathbf{x}}_i, t_i} &~ t_i \nonumber\\
\text{s.t.} &~ \Re\left\{\hat{\mathbf{h}}^T_{k}\tilde{\mathbf{x}}_i\right\}-\frac{\left|\Im\left\{\hat{\mathbf{h}}^T_{k}\tilde{\mathbf{x}}_i\right\}\right|}{\tan \frac{\pi}{\mathcal{M}}}\geq t_i \sqrt{\gamma_{k,i}}\sigma,\, \forall k,\nonumber\\
&~ \left\|\tilde{\mathbf{x}}_i\right\|^2\leq p_i.
\end{align}
Before investigating the block-level SB slot-variant IE precoder's relationship to SB-SLP, let us review the explicit duality for PM/SB-SLP.

\begin{lemma}[Duality for Symbol-Level IE Precoding \cite{10229497}]
\label{lemmaSLDuality}
\it The PM-SLP problem \eqref{eqSLPMx} and the SB-SLP problem \eqref{eqSLSB} are dual problems. Let $\mathbf{x}^{PM}_{i}\left(\left\{\gamma_{k,i}\right\}\right)$ and $p^{PM}_{i}\left(\left\{\gamma_{k,i}\right\}\right)\triangleq\|\mathbf{x}^{PM}_{i}\left(\left\{\gamma_{k,i}\right\}\right)\|^2$ denote the optimal solution and the optimal value of the PM-SLP problem (\ref{eqSLPMx}) given $\left\{\gamma_{k,i}\right\}$, respectively. Then the counterparts for the SB-SLP problem (\ref{eqSLSB}), $\mathbf{x}^{SB}_{i}\left(\left\{\gamma_{k,i}\right\},p_i\right)$ and $\mu^{SB}_{i}\left(\left\{\gamma_{k,i}\right\},p_i\right)$, are given by
\begin{align}
\tilde{\mathbf{x}}^{SB}_{i}\left(\left\{\gamma_{k,i}\right\},p_i\right)=&\sqrt{\frac{p_i}{p^{PM}_{i}\left(\left\{\gamma_{k,i}\right\}\right)}}\tilde{\mathbf{x}}^{PM}_{i}\left(\left\{\gamma_{k,i}\right\}\right),\\
t^{SB}_{i}\left(\left\{\gamma_{k,i}\right\},p_{i}\right)=&\sqrt{\frac{p_i}{p^{PM}_{i}\left(\left\{\gamma_{k,i}\right\}\right)}}.
\end{align}
and vice versa as
\begin{align}
\tilde{\mathbf{x}}^{PM}_{i}\left(\left\{\gamma_{k,i}\right\}\right)=&\frac{1}{t^{SB}_{i}\left(\left\{\gamma_{k,i}\right\},p_{i}\right)}\tilde{\mathbf{x}}^{SB}_{i}\left(\left\{\gamma_{k,i}\right\},p_i\right),\\
p^{PM}_{i}\left(\left\{\gamma_{k,i}\right\}\right)=&\frac{p_{i}}{\left(t^{SB}_{i}\left(\left\{\gamma_{k,i}\right\},p_{i}\right)\right)^2}.
\end{align}
\end{lemma}

\begin{IEEEproof}[\bf Proof]
A detailed proof can be found in \cite{10229497}.
\end{IEEEproof}

A useful explicit duality for the block-level slot-variant IE precoders can be subsequently demonstrated below. 

\begin{theorem}[Duality for Block-Level Slot-Variant IE Precoding]
\label{theorBLDuality}
\it The block-level PM slot-variant IE precoder in \eqref{eqBLPMx} and the SB slot-variant IE precoder in \eqref{eqBLSBcomplex} are dual problems. Let $\left\{\tilde{\mathbf{x}}^{PM}_{i}\left(\left\{\gamma_{k,i}\right\}\right)\right\}$ and $\sum^{N_s}_{i=1}p^{PM}_{i}\left(\left\{\gamma_{k,i}\right\}\right)\triangleq\sum^{N_s}_{i=1}\|\tilde{\mathbf{x}}^{PM}_{i}\left(\left\{\gamma_{k,i}\right\}\right)\|^2$ denote the optimal solution and the optimal value of the block-level PM slot-variant IE precoding problem (\ref{eqBLPMx}) given $\left\{\gamma_{k,i}\right\}$, respectively. Then the counterparts for the block-level SB slot-variant IE precoding problem (\ref{eqBLSBcomplex}), $\left\{\tilde{\mathbf{x}}^{SB}_{i}\left(\left\{\gamma_{k,i}\right\},\sum^{N_s}_{i=1}p_i\right)\right\}$ and $t^{SB}\left(\left\{\gamma_{k,i}\right\},\sum^{N_s}_{i=1}p_i\right)$, are determined as
\begin{align}
\tilde{\mathbf{x}}^{SB}_{i}\left(\left\{\gamma_{k,i}\right\},\sum^{N_s}_{i=1}p_i\right)=&\sqrt{\frac{\sum^{N_s}_{i=1}p_i}{\sum^{N_s}_{i=1}p^{PM}_{i}\left(\left\{\gamma_{k,i}\right\}\right)}}\times\nonumber\\
&\tilde{\mathbf{x}}^{PM}_{i}\left(\left\{\gamma_{k,i}\right\}\right),\forall i,\\
t^{SB}\left(\left\{\gamma_{k,i}\right\},\sum^{N_s}_{i=1}p_{i}\right)=&\sqrt{\frac{\sum^{N_s}_{i=1}p_i}{\sum^{N_s}_{i=1}p^{PM}_{i}\left(\left\{\gamma_{k,i}\right\}\right)}}.
\end{align}
and vice versa as
\begin{align}
\tilde{\mathbf{x}}^{PM}_{i}\left(\left\{\gamma_{k,i}\right\}\right)=&\frac{1}{t^{SB}\left(\left\{\gamma_{k,i}\right\},\sum^{N_s}_{i=1}p_{i}\right)}\times\nonumber\\
&\tilde{\mathbf{x}}^{SB}_{i}\left(\left\{\gamma_{k,i}\right\},\sum^{N_s}_{i=1}p_i\right),\forall i,\\
\sum^{N_s}_{i=1}p^{PM}_{i}\left(\left\{\gamma_{k,i}\right\}\right)=&\frac{\sum^{N_s}_{i=1}p_{i}}{\left(t^{SB}\left(\left\{\gamma_{k,i}\right\},\sum^{N_s}_{i=1}p_{i}\right)\right)^2}.
\end{align}
\end{theorem}
\begin{IEEEproof}[\bf Proof]
The proof is similar to that of Lemma \ref{lemmaSLDuality} and is therefore omitted.
\end{IEEEproof}

\begin{theorem}
\label{theorEquaSoluSB}
\it The optimal solutions to the block-level SB slot-variant IE precoder in \eqref{eqBLSBcomplex} and the SB-SLP problem \eqref{eqSLSB} are symbol-level scaled versions of each other. Let $\tilde{\mathbf{x}}^{BL}_i$ and $\tilde{\mathbf{x}}^{SL}_i$ be the solutions in the $i$-th symbol slot to the block-level SB slot-variant IE precoding problem \eqref{eqBLSBcomplex} and the SB-SLP problem \eqref{eqSLSB}, respectively. These solutions are related as follows:
\begin{align}
\label{eqAllocBLSL}
\tilde{\mathbf{x}}^{BL}_i=&\frac{1}{t_i}\sqrt{\frac{\sum_{j=1}^{N_s}p_j}{\sum_{j=1}^{N_s}\frac{p_j}{t_j^2}}}\tilde{\mathbf{x}}^{SL}_i,\forall i,\\
\tilde{\mathbf{x}}^{SL}_i=&\sqrt{\frac{p_i}{\left\|\tilde{\mathbf{x}}^{BL}_i\right\|^2}}\tilde{\mathbf{x}}^{BL}_i,\forall i.
\label{eqAllocSLBL}
\end{align}
\end{theorem}
\begin{IEEEproof}[\bf Proof]
From Theorem \ref{theorBLDuality}, we see that the block-level SB slot-variant IE precoder in \eqref{eqBLSBcomplex} is a scaled version of the block-level PM slot-variant IE precoder in \eqref{eqBLPMx}. Together with Theorem \ref{theorEquaSolu}, it follows that the block-level SB slot-variant IE precoder in \eqref{eqBLSBcomplex} is a power scaled version of the optimal solution to the PM-SLP problem \eqref{eqSLPMx}. From Lemma \ref{lemmaSLDuality}, we further conclude that the block-level SB slot-variant IE precoder in \eqref{eqBLSBcomplex} is a symbol-level scaled version of the optimal solution to the SB-SLP problem \eqref{eqSLSB}, and vice versa. The rest of this proof addresses \eqref{eqAllocBLSL} and \eqref{eqAllocSLBL}.

Based on Lemma \ref{lemmaSLDuality}, the optimal transmit power of the PM-SLP problem is $p^{PM}_i=p_i$, and the optimal auxiliary variable of the SB-SLP problem is $t^{SB}_i=1$. Therefore the minimum instantaneous received SINRs in each symbol slot $\left\{t^{SB}_i\right\}$ are identical and equal to $1$ in this case.

Given an arbitrary block-level transmit power budget, the optimal solution to the block-level SB slot-variant IE precoding problem \eqref{eqBLSBcomplex} can be derived from the optimal solution to the SB-SLP problem \eqref{eqSLSB} by normalizing the minimum instantaneous received SINRs to the same level, i.e., multiplying by $\frac{1}{t_i}$, and subsequently rescaling the transmit power to satisfy the block-level transmit power constraint by multiplying by $\sqrt{\frac{\sum_{j=1}^{N_s}p_j}{\sum_{j=1}^{N_s}\frac{p_j}{t_j^2}}}$. Conversely, the optimal solution to the SB-SLP problem \eqref{eqSLSB} can be obtained by rescaling the block-level SB slot-variant IE precoder such that it satisfies the symbol-level transmit power constraint.
\end{IEEEproof}

\section{Block-Level Slot-Invariant Interference-Exploitation Precoder}
\label{secSIP}
To find a further compromise between performance and complexity, we design block-level PM and SB slot-invariant IE precoders in this section.

\subsection{PM Slot-Invariant IE Precoder}
The goal of the PM slot-invariant IE precoder is to minimize the block-level transmit power while adhering to CI constraints. In comparison to the PM slot-variant IE precoder design, this approach involves shifting the optimization variable from slot-variant IE precoders to a unified slot-invariant IE precoder. The optimization problem for the PM slot-invariant IE precoder is given by
\begin{align}
\label{eqBLPMSI}
\min_{\tilde{\mathbf{W}}} &~ \sum_{i=1}^{N_s}\left\|\tilde{\mathbf{W}}\tilde{\mathbf{s}}_{:,i}\right\|^2 \nonumber\\
\text{s.t.} &~ \Re\left\{\hat{\mathbf{h}}^T_{k}\tilde{\mathbf{W}}\tilde{\mathbf{s}}_{:,i}\right\}-\frac{\left|\Im\left\{\hat{\mathbf{h}}^T_{k}\tilde{\mathbf{W}}\tilde{\mathbf{s}}_{:,i}\right\}\right|}{\tan \frac{\pi}{\mathcal{M}}}\geq\sqrt{\gamma_{k,i}}\sigma,\, \forall k,\forall i,
\end{align}
This problem is a linear constrained quadratic programming, and thus can be solved by off-the-shelf convex optimizers.

To rewrite the complex-valued problem in \eqref{eqBLPMSI}, we introduce the following set of real-valued matrices:
\begin{subequations}
\label{eqWstruc}
\begin{align}
\mathbf{W}\triangleq &\begin{bmatrix}
\Re\left\{\tilde{\mathbf{W}}\right\}&-\Im\left\{\tilde{\mathbf{W}}\right\}\\
\Im\left\{\tilde{\mathbf{W}}\right\}&\Re\left\{\tilde{\mathbf{W}}\right\}
\end{bmatrix}\in \mathbb{R}^{2N_t\times 2N_r},\\
\overline{\mathbf{W}}
\triangleq &\begin{bmatrix}
\Re\left\{\tilde{\mathbf{W}}\right\}&-\Im\left\{\tilde{\mathbf{W}}\right\}\\
\end{bmatrix}\in \mathbb{R}^{N_t \times 2N_r},\\
\mathbf{P}_1 \triangleq &\begin{bmatrix}
\mathbf{I}\\ \mathbf{0}
\end{bmatrix}\in \mathbb{R}^{2N_t \times N_t},\\
\mathbf{P}_2 \triangleq &\begin{bmatrix}
\mathbf{0}\\ \mathbf{I}
\end{bmatrix}\in \mathbb{R}^{2N_t \times N_t},\\
\mathbf{P}_3\triangleq &\begin{bmatrix}
\mathbf{0}&\mathbf{I}\\
-\mathbf{I}&\mathbf{0}
\end{bmatrix}\in \mathbb{R}^{2N_r\times 2N_r}.
\end{align}
\end{subequations}
It is easy to verify that 
\begin{align}
\mathbf{W}=&\mathbf{P}_1\overline{\mathbf{W}}+\mathbf{P}_2\overline{\mathbf{W}}\mathbf{P}_3,\\
\mathbf{P}^T_1\mathbf{P}_1=&\mathbf{P}^T_2\mathbf{P}_2=\mathbf{I},\\
\mathbf{P}^T_1\mathbf{P}_2=&\mathbf{P}^T_2\mathbf{P}_1=\mathbf{0}.
\end{align}
We are now prepared to formulate the real-valued problem that corresponds to the complex-valued problem for the PM slot-invariant IE precoder \eqref{eqBLPMSI}:
\begin{align}
\label{eqPMSIre}
\min_{\mathbf{W},\overline{\mathbf{W}}} &~ \sum_{i=1}^{N_s}\left\|\mathbf{W}\mathbf{s}_{:,i}\right\|^2 \nonumber\\
\text{s.t.} &~ \mathbf{T}\hat{\mathbf{S}}_i\mathbf{H}\mathbf{W}\mathbf{s}_{:,i}\succeq\mathbf{b}_i,\forall i, ~
~ \mathbf{W}=\mathbf{P}_1\overline{\mathbf{W}}+\mathbf{P}_2\overline{\mathbf{W}}\mathbf{P}_3.
\end{align}
The matrix constraint $\mathbf{W}=\mathbf{P}_1\overline{\mathbf{W}}+\mathbf{P}_2\overline{\mathbf{W}}\mathbf{P}_3$ is introduced to ensure that the optimal real-valued precoder $\mathbf{W}$ arranges the real and imaginary parts of the complex-valued precoder $\tilde{\mathbf{W}}$ as defined in \eqref{eqWstruc}. This problem can be simplified by eliminating the variable matrix $\mathbf{W}$:
\begin{align}
\label{eqSLPMre}
\min_{\overline{\mathbf{W}}} &~ \sum_{i=1}^{N_s}\left\|\left(\mathbf{P}_1\overline{\mathbf{W}}+\mathbf{P}_2\overline{\mathbf{W}}\mathbf{P}_3\right)\mathbf{s}_{:,i}\right\|^2 \nonumber\\
\text{s.t.} &~ \mathbf{T}\hat{\mathbf{S}}_i\mathbf{H}\left(\mathbf{P}_1\overline{\mathbf{W}}+\mathbf{P}_2\overline{\mathbf{W}}\mathbf{P}_3\right)\mathbf{s}_{:,i}\succeq\mathbf{b}_i,\forall i.
\end{align}
It can be seen that the above problem preserves the quadratic objective function and linear constraints of \eqref{eqBLPMSI}. However, directly handling this problem is complicated due to the high-dimensional optimization variable $\overline{\mathbf{W}}$ and the underlying matrix structure constraint. To address this, we will examine the problem from a Lagrangian dual perspective in the following subsections.

\subsection{Optimal Structure for PM Slot-Invariant IE Precoder}
\label{secStrucPMSI}
To formulate the Lagrangian dual problem, we begin by writing the Lagrangian function associated with \eqref{eqSLPMre} as follows:
\begin{align}
\mathcal{L}\left(\overline{\mathbf{W}},\right\{\boldsymbol{\lambda}_i\left\}\right)\triangleq  \sum_{i=1}^{N_s}\left\|\left(\mathbf{P}_1\overline{\mathbf{W}}+\mathbf{P}_2\overline{\mathbf{W}}\mathbf{P}_3\right)\mathbf{s}_{:,i}\right\|^2\nonumber\\
+\boldsymbol{\lambda}^T_i\left[\mathbf{b}_i-\mathbf{T}\hat{\mathbf{S}}_i\mathbf{H}\left(\mathbf{P}_1\overline{\mathbf{W}}+\mathbf{P}_2\overline{\mathbf{W}}\mathbf{P}_3\right)\mathbf{s}_{:,i}\right],
\end{align}
where $\boldsymbol{\lambda}_i$ is the non-negative Lagrange dual variable vector or Lagrange multiplier associated with the CI constraints $\mathbf{T}\hat{\mathbf{S}}_i\mathbf{H}\left(\mathbf{P}_1\overline{\mathbf{W}}+\mathbf{P}_2\overline{\mathbf{W}}\mathbf{P}_3\right)\mathbf{s}_{:,i}\succeq\mathbf{b}_i$ in the $i$-th symbol slot. From the convexity of \eqref{eqSLPMre}, the duality gap is zero. When the primal variable $\overline{\mathbf{W}}$ and dual variable $\left\{\boldsymbol{\lambda}_i\right\}$ achieve optimality, the following KKT conditions must be satisfied:
\begin{align}
\mathbf{T}\hat{\mathbf{S}}_i\mathbf{H}\left(\mathbf{P}_1\overline{\mathbf{W}}+\mathbf{P}_2\overline{\mathbf{W}}\mathbf{P}_3\right)\mathbf{s}_{:,i}\succeq &\mathbf{b}_i,\forall i, \\
\boldsymbol{\lambda}_i\succeq &\mathbf{0},\forall i, \\
\boldsymbol{\lambda}_{i,k}\left[\mathbf{T}\hat{\mathbf{S}}_i\mathbf{H}\left(\mathbf{P}_1\overline{\mathbf{W}}+\mathbf{P}_2\overline{\mathbf{W}}\mathbf{P}_3\right)\right]_{k,:}\mathbf{s}_{:,i}=&\boldsymbol{\lambda}_{i,k}\mathbf{b}_{i,k}, \forall k, \forall i,\\
\label{eqStationarity}
\frac{\partial\mathcal{L}\left(\overline{\mathbf{W}},\right\{\boldsymbol{\lambda}_i\left\}\right)}{\partial\overline{\mathbf{W}}}=&\mathbf{0}, 
\end{align}
where $\boldsymbol{\lambda}_{i,k}$ and $\left[\mathbf{T}\hat{\mathbf{S}}_i\mathbf{H}\left(\mathbf{P}_1\overline{\mathbf{W}}+\mathbf{P}_2\overline{\mathbf{W}}\mathbf{P}_3\right)\right]_{k,:}$ respectively denote the $k$-th component of $\boldsymbol{\lambda}_i$ and the $k$-th row of $\mathbf{T}\hat{\mathbf{S}}_i\mathbf{H}\left(\mathbf{P}_1\overline{\mathbf{W}}+\mathbf{P}_2\overline{\mathbf{W}}\mathbf{P}_3\right)$.

The stationarity condition in \eqref{eqStationarity} can be attained by setting the partial derivative of $\mathcal{L}\left(\overline{\mathbf{W}},\right\{\boldsymbol{\lambda}_i\left\}\right)$ with respect to the primal variable $\overline{\mathbf{W}}$ to zero:
\begin{align}
\sum_{i=1}^{N_s}2\overline{\mathbf{W}}\left(\mathbf{s}_{:,i}\mathbf{s}^T_{:,i}+\mathbf{P}_3\mathbf{s}_{:,i}\mathbf{s}^T_{:,i}\mathbf{P}^T_3\right)-\mathbf{P}^T_1\mathbf{H}^T\hat{\mathbf{S}}^T_i\mathbf{T}^T\boldsymbol{\lambda}_i\mathbf{s}^T_{:,i}\nonumber\\
-\mathbf{P}^T_2\mathbf{H}^T\hat{\mathbf{S}}^T_i\mathbf{T}^T\boldsymbol{\lambda}_i\mathbf{s}^T_{:,i}\mathbf{P}^T_3=\mathbf{0}.
\end{align}
When $N_s\geq N_r$, the rank of matrix $\sum_{i=1}^{N_s}\mathbf{s}_{:,i}\mathbf{s}^T_{:,i}+\mathbf{P}_3\mathbf{s}_{:,i}\mathbf{s}^T_{:,i}\mathbf{P}^T_3$ is $2N_r$, indicating that it is non-singular. With this information, the optimal solution of the real-valued PM problem \eqref{eqSLPMre} is given by
\begin{align}
\label{eqSoluStruc}
\overline{\mathbf{W}}=\frac{\sum_{i=1}^{N_s}\mathbf{P}^T_1\mathbf{H}^T\hat{\mathbf{S}}^T_i\mathbf{T}^T\boldsymbol{\lambda}_i\mathbf{s}^T_{:,i}+\mathbf{P}^T_2\mathbf{H}^T\hat{\mathbf{S}}^T_i\mathbf{T}^T\boldsymbol{\lambda}_i\mathbf{s}^T_{:,i}\mathbf{P}^T_3}{2\sum_{i=1}^{N_s}\mathbf{s}_{:,i}\mathbf{s}^T_{:,i}+\mathbf{P}_3\mathbf{s}_{:,i}\mathbf{s}^T_{:,i}\mathbf{P}^T_3}.
\end{align}

The KKT conditions indicate that the optimal primal variable can optimize the Lagrangian function $\mathcal{L}\left(\overline{\mathbf{W}},\right\{\boldsymbol{\lambda}_i\left\}\right)$ at the zero derivative point. Having obtained the optimal solution in \eqref{eqSoluStruc}, we can substitute it into the Lagrangian function to eliminate the primal variable $\overline{\mathbf{W}}$. This process allows us to obtain the dual function, as given by Proposition \ref{propDualFunc} below. 
\begin{proposition}
\label{propDualFunc}
\it Assuming that $N_s\geq N_r$, the dual function of the problem for the PM slot-invariant IE precoder \eqref{eqSLPMre} is given by
\begin{align}
\label{eqDualFunc}
g\left(\boldsymbol{\lambda}\right)\triangleq &-\frac{1}{4}\boldsymbol{\lambda}^T\mathbf{U}\boldsymbol{\lambda}+\boldsymbol{\lambda}^T\mathbf{b}.
\end{align}
\end{proposition}
\begin{IEEEproof}[\bf Proof]
See Appendix \ref{proofDualFunc}.
\end{IEEEproof}

As a result, we obtain the dual problem for the PM slot-invariant IE precoder \eqref{eqSLPMre} as follows:
\begin{align}
\label{eqPMSIdual}
\min_{\boldsymbol{\lambda}} ~  \frac{1}{4}\boldsymbol{\lambda}^T\mathbf{U}\boldsymbol{\lambda}-\mathbf{b}^T\boldsymbol{\lambda} ~
\text{s.t.} ~ \boldsymbol{\lambda}\succeq\mathbf{0}.
\end{align}
Compared with the primal problem in \eqref{eqSLPMre}, the structure of the dual problem in \eqref{eqPMSIdual} is much simpler. First, the two problems have different optimization variables. By deriving the dual problem, we reduce the dimension of the optimization variable from $2N_s N_t \times 2N_s N_r$ to $2N_s N_r$. Second, although both problems are linearly constrained quadratic programming problems, the non-negative constraints in the dual problem are easier to handle compared to the polyhedral constraints in the primal problem.

In the scenario where $N_s<N_r$, the rank of matrix $\sum_{i=1}^{N_s}\mathbf{s}_{:,i}\mathbf{s}^T_{:,i}+\mathbf{P}_3\mathbf{s}_{:,i}\mathbf{s}^T_{:,i}\mathbf{P}^T_3$ is $2N_s$ and thus it is singular. A unique Moore-Penrose pseudoinverse can be used to replace the matrix inverse of $\sum_{i=1}^{N_s}\mathbf{s}_{:,i}\mathbf{s}^T_{:,i}+\mathbf{P}_3\mathbf{s}_{:,i}\mathbf{s}^T_{:,i}\mathbf{P}^T_3$ in the above optimal structure and dual problem.

\subsection{SB Slot-Invariant IE Precoder}
Following the previous designs for block-level slot-variant and slot-invariant IE precoders, the slot-invariant IE precoder proposed in this section can also be applied to the weighted SB problem. As a result, the complex-valued optimization problem for the weighted SB slot-invariant IE precoder can be represented as
\begin{align}
\label{eqBLSBSIcomplex}
\max_{\tilde{\mathbf{W}},t} &~ t \nonumber\\
\text{s.t.}\,& \Re\left\{\hat{\mathbf{h}}^T_{k}\tilde{\mathbf{W}}\tilde{\mathbf{s}}_{:,i}\right\}-\frac{\left|\Im\left\{\hat{\mathbf{h}}^T_{k}\tilde{\mathbf{W}}\tilde{\mathbf{s}}_{:,i}\right\}\right|}{\tan \frac{\pi}{\mathcal{M}}}\geq t \sqrt{\gamma_{k,i}}\sigma,\, \forall k, \forall i, \nonumber\\
\, &\sum_{i=1}^{N_s}\left\|\tilde{\mathbf{W}}\tilde{\mathbf{s}}_{:,i}\right\|^2\leq \sum_{i=1}^{N_s}p_i.
\end{align}

The real-valued form of \eqref{eqBLSBSIcomplex} can be written as
\begin{align}
\label{eqBLSBre}
\max_{\overline{\mathbf{W}},t} &~ t \nonumber\\
\text{s.t.} &~ \mathbf{T}\hat{\mathbf{S}}_{i}\mathbf{H}\left(\mathbf{P}_1\overline{\mathbf{W}}+\mathbf{P}_2\overline{\mathbf{W}}\mathbf{P}_3\right)\mathbf{s}_{:,i}\geq t\mathbf{b}_i,\, \forall i, \nonumber\\
&~ \sum_{i=1}^{N_s}\left\|\left(\mathbf{P}_1\overline{\mathbf{W}}+\mathbf{P}_2\overline{\mathbf{W}}\mathbf{P}_3\right)\mathbf{s}_{:,i}\right\|^2\leq \sum_{i=1}^{N_s}p_i.
\end{align}
Similar to Proposition \ref{PropActive}, the above block-level transmit power constraint is also active as long as optimality is achieved.

When $\left\{\sqrt{\gamma_{k,i}}\sigma\right\}$ equals 1, this problem is equivalent to the optimization problem proposed in \cite{9962829}, referred to as CI-BLP, in which the \emph{symbol-scaling} CI metric was employed to formulate the problem that maximizes the minimum CI effect of an entire transmission block and subject to a block-level power constraint.

\subsection{Optimal Structure for SB Slot-Invariant IE Precoder}
For the sake of completeness, we give the optimal solution of the real-valued problem for SB slot-invariant IE precoder below and omit the derivations:
\begin{equation}
\label{eqSoluStrucSB}
\overline{\mathbf{W}}=\frac{\sum_{i=1}^{N_s}\mathbf{P}^T_1\mathbf{H}^T\hat{\mathbf{S}}^T_i\mathbf{T}^T\boldsymbol{\lambda}_i\mathbf{s}^T_{:,i}+\mathbf{P}^T_2\mathbf{H}^T\hat{\mathbf{S}}^T_i\mathbf{T}^T\boldsymbol{\lambda}_i\mathbf{s}^T_{:,i}\mathbf{P}^T_3}{2\mu\sum_{i=1}^{N_s}\mathbf{s}_{:,i}\mathbf{s}^T_{:,i}+\mathbf{P}_3\mathbf{s}_{:,i}\mathbf{s}^T_{:,i}\mathbf{P}^T_3},
\end{equation}
where
$
\mu=\frac{1}{2}\sqrt{\frac{\boldsymbol{\lambda}^T\mathbf{U}\boldsymbol{\lambda}}{\sum_{i=1}^{N_s}p_i}}.
$

Following a procedure similar to that in Section \ref{secStrucSBSV} and Section \ref{secStrucPMSI}, the dual problem of \eqref{eqBLSBre} can be shown to be
\begin{align}
\label{eqSBdual}
\min_{\boldsymbol{\lambda}}~ \boldsymbol{\lambda}^T\mathbf{U}\boldsymbol{\lambda} ~
\text{s.t.}~ \boldsymbol{\lambda}\succeq\mathbf{0},
~ \mathbf{b}^T\boldsymbol{\lambda}-1=0.
\end{align}

\subsection{Duality Between PM and SB Slot-Invariant IE Precoders}
\label{secDualSI}
In this subsection, we investigate the properties of the PM and SB slot-invariant IE precoding problems, and extend the explicit duality for block-level slot-variant IE precoding proposed in Section \ref{secRela} to slot-invariant IE precoding.

Let $\tilde{\mathbf{W}}^{PM}\left(\left\{\gamma_{k,i}\right\}\right)$ and $\sum_{i=1}^{N_s}p^{PM}_{i}\left(\left\{\gamma_{k,i}\right\}\right)\triangleq\sum_{i=1}^{N_s}\left\|\tilde{\mathbf{W}}^{PM}\left(\left\{\gamma_{k,i}\right\}\right)\tilde{\mathbf{s}}_{:,i}\right\|^2$ denote the optimal solution and objective value of the block-level PM slot-invariant IE precoding problem \eqref{eqBLPMSI} given $\left\{\gamma_{k,i}\right\}$. $\tilde{\mathbf{W}}^{SB}\left(\left\{\gamma_{k,i}\right\}, \sum_{i=1}^{N_s}p_i\right)$ and $t^{SB}\left(\left\{\gamma_{k,i}\right\}, \sum_{i=1}^{N_s}p_i\right)$ are the optimal counterparts for the block-level SB slot-invariant IE precoding problem \eqref{eqBLSBSIcomplex} given $\left\{\gamma_{k,i}\right\}$ and $\sum_{i=1}^{N_s}p_i$.
\begin{theorem}[Duality for Block-Level Slot-Invariant IE Precoding]
\label{theorBLDualitySI}
\it The block-level PM slot-invariant IE precoder in \eqref{eqBLPMSI} and SB slot-invariant IE precoder in \eqref{eqBLSBSIcomplex} are dual problems. The explicit duality between them can be expressed as
\begin{align}
\tilde{\mathbf{W}}^{SB}\left(\left\{\gamma_{k,i}\right\},\sum^{N_s}_{i=1}p_i\right)=&\sqrt{\frac{\sum^{N_s}_{i=1}p_i}{\sum^{N_s}_{i=1}p^{PM}_{i}\left(\left\{\gamma_{k,i}\right\}\right)}}\times\nonumber\\
& \tilde{\mathbf{W}}^{PM}\left(\left\{\gamma_{k,i}\right\}\right),\forall i,\\
t^{SB}\left(\left\{\gamma_{k,i}\right\},\sum^{N_s}_{i=1}p_{i}\right)=&\sqrt{\frac{\sum^{N_s}_{i=1}p_i}{\sum^{N_s}_{i=1}p^{PM}_{i}\left(\left\{\gamma_{k,i}\right\}\right)}}.
\end{align}
and vice versa as
\begin{align}
\tilde{\mathbf{W}}^{PM}\left(\left\{\gamma_{k,i}\right\}\right)=&\frac{1}{t^{SB}\left(\left\{\gamma_{k,i}\right\},\sum^{N_s}_{i=1}p_{i}\right)}\times\nonumber\\
&\tilde{\mathbf{W}}^{SB}\left(\left\{\gamma_{k,i}\right\},\sum^{N_s}_{i=1}p_i\right),\forall i,\\
\sum^{N_s}_{i=1}p^{PM}_{i}\left(\left\{\gamma_{k,i}\right\}\right)=&\frac{\sum^{N_s}_{i=1}p_{i}}{\left(t^{SB}\left(\left\{\gamma_{k,i}\right\},\sum^{N_s}_{i=1}p_{i}\right)\right)^2}.
\end{align}
\end{theorem}
\begin{IEEEproof}[\bf Proof]
Verbatim to the proof of Theorem 1 in \cite{10229497}.
\end{IEEEproof}

\section{Linear Precoder for Interference Exploitation}
\label{secLinear}
In this section, we present an in-depth investigation on the existence of a linear precoder for IE. Additionally, we demonstrate the intrinsic connections between the block-level precoders proposed in this paper and the conventional linear precoder. Furthermore, we propose a novel linear precoder under the assumption that the number of symbol slots does not exceed the number of users. Under this assumption, we can equivalently decompose the slot-invariant IE precoding problem into subproblems over each symbol slot.

We also highlight that the linear precoder can be used to estimate the data symbols in a variety of practical multi-antenna systems. A notable example is the QR-maximum likelihood detector (QR-MLD), which estimates the data symbols at the receiver side by performing QR decomposition of the product of the channel matrix and the linear precoder \cite{1237128,7244171}.

When SLP or BLP is adopted to design the transmit signal, all the $N_r\times N_s$ received signals within one transmission block can be compactly aggregated in one matrix, which can be written as
\begin{align}
\tilde{\mathbf{Y}}=\tilde{\mathbf{H}}\tilde{\mathbf{W}}\tilde{\mathbf{S}}+\tilde{\mathbf{N}}=\tilde{\mathbf{H}}\tilde{\mathbf{X}}+\tilde{\mathbf{N}},
\end{align}
where 
\begin{subequations}
\begin{align}
\tilde{\mathbf{Y}}\triangleq &
\begin{bmatrix}
\tilde{y}_{1,1}&\cdots&\tilde{y}_{1,N_s}\\
\vdots&\ddots&\vdots\\
\tilde{y}_{N_r,1}&\cdots&\tilde{y}_{N_r,N_s}
\end{bmatrix}\in \mathbb{C}^{N_r \times N_s},\\
\tilde{\mathbf{H}}\triangleq &
\begin{bmatrix}
\tilde{\mathbf{h}}_1&\cdots&\tilde{\mathbf{h}}_{N_r}
\end{bmatrix} \in \mathbb{C}^{N_r \times N_t},\\
\tilde{\mathbf{S}}\triangleq &
\begin{bmatrix}
\tilde{\mathbf{s}}_{:,1}&\cdots&\tilde{\mathbf{s}}_{:,N_s}
\end{bmatrix}\in \mathbb{C}^{N_r \times N_s},\\
\tilde{\mathbf{N}}\triangleq &
\begin{bmatrix}
\tilde{n}_{1,1}&\cdots&\tilde{n}_{1,N_s}\\
\vdots&\ddots&\vdots\\
\tilde{n}_{N_r,1}&\cdots&\tilde{n}_{N_r,N_s}
\end{bmatrix}\in \mathbb{C}^{N_r \times N_s},\\
\tilde{\mathbf{X}}\triangleq &
\begin{bmatrix}
\tilde{\mathbf{x}}_1&\cdots&\tilde{\mathbf{x}}_{N_s}
\end{bmatrix}\in \mathbb{C}^{N_t \times N_s}.
\end{align}
\end{subequations}
When $N_{s} \le N_{r}$, we can represent the optimal transmit signal matrix as a linear transformation of the data symbols:
\begin{equation}
\label{eqLinearSys}
\tilde{\mathbf{X}}=\tilde{\mathbf{W}}\tilde{\mathbf{S}},
\end{equation}
where
\begin{equation}
\label{eqWxs}
\tilde{\mathbf{W}}=\tilde{\mathbf{X}}\tilde{\mathbf{S}}^{\dagger}.
\end{equation}
If $\tilde{\mathbf{S}}$ has full column rank, we have
\begin{equation} 
\tilde{\mathbf{S}}^{\dagger}=\left(\tilde{\mathbf{S}}^{H}\tilde{\mathbf{S}}\right)^{-1}\tilde{\mathbf{S}}^{H}. 
\end{equation}
On the other hand, if $\tilde{\mathbf{S}}$ is rank-deficient, we can replace $\tilde{\mathbf{S}}^{\dagger}$ by the Moore-Penrose pseudoinverse matrix of $\tilde{\mathbf{S}}$, which can be readily computed using the singular value decomposition of $\tilde{\mathbf{S}}$.

The precoder in \eqref{eqWxs} can be regarded as a unified linear transformation matrix applied to the data symbol vectors $\left\{\tilde{\mathbf{s}}_{:,i}\right\}$ over the $N_s $ symbol slots. In each symbol slot, the precoded signal or transmit signal can be expressed as
\begin{equation}
\tilde{\mathbf{x}}_i=\tilde{\mathbf{W}}\tilde{\mathbf{s}}_{:,i}, \forall i.
\end{equation}
This linear precoder holds as a general solution, regardless of the construction of the transmit signal matrix $\tilde{\mathbf{X}}$. By extending the above linear precoder structure to IE precoding, we establish the following theorem.
\begin{theorem}
\label{theorLinear}
\it Let $N_s\leq N_r$, then there exists a linear precoder for IE precoding, given by
\begin{align}
\tilde{\mathbf{W}}=\tilde{\mathbf{X}}\tilde{\mathbf{S}}^{\dagger}=\begin{bmatrix}
\tilde{\mathbf{W}}_1\tilde{\mathbf{s}}_{:,1}&\cdots&\tilde{\mathbf{W}}_{N_s}\tilde{\mathbf{s}}_{:,N_s}
\end{bmatrix}\tilde{\mathbf{S}}^{\dagger},
\end{align}
which guarantees a constant precoder for each symbol slot within one transmission block.
\end{theorem} 

Accordingly, for symbol-level IE precoding, we can first solve the $N_s \leq N_r$ SLP problems and acquire the $N_s$ optimal symbol-level precoders $\left\{\tilde{\mathbf{W}}_i\right\}$ or equivalently the $N_s$ optimal symbol-level transmit signal vectors $\left\{\tilde{\mathbf{x}}_i\right\}$, then construct the linear precoder based on Theorem \ref{theorLinear}.

\begin{remark}
\it When the number of symbol slots does not exceed the number of users, i.e., $N_s\leq N_r$, the block-level slot-variant IE precoding and slot-invariant IE precoding have an identical precoder or transmit signal in each symbol slot, which means that the matrix structure constraint in \eqref{eqPMSIre} is redundant and can be discarded. Based on Section \ref{secSVP} and Section \ref{secSIP}, the slot-variant IE precoding can be decomposed into $N_s$ smaller subproblems over $N_s$ symbol slots, whereas the slot-invariant IE precoding cannot due to the extra structural constraint in \eqref{eqPMSIre}. Therefore, Theorem \ref{theorLinear} indicates that under the assumption of $N_s\leq N_r$, we can address the simpler slot-variant IE precoding problem rather than the slot-invariant IE precoding problem to obtain a linear or slot-invariant IE precoder.
\end{remark}

\section{Numerical Results}
\label{secResults}
In this section, numerical results based on Monte Carlo simulations are presented to evaluate the proposed block-level slot-variant and slot-invariant IE precoders, as well as the linear precoder for IE. For the simulations of the PM problem, we assume each user has an identical SINR threshold and unit noise variance, i.e., $\gamma_{k,i}=\gamma, \forall k, \forall i$, and $\sigma^2=1$. For the SB problem, the weights of the users' SINR are assumed to be $\gamma_{k,i}=\gamma=1, \forall k, \forall i$. In each symbol slot, the SNR is determined as $\text{SNR}=\frac{p_i}{\sigma^2}=\frac{1}{\sigma^2}, \forall i$.  The block-level transmit power budget is set to $\sum^{N_s}_{i=1}p_i=N_s$. The system setting $N_r \times N_t $ is indicated in the legend of each figure.

For clarity, we list the abbreviations of the considered algorithms below.

In simulations for PM problems:
\begin{enumerate}
\item `ZF': Linear ZF precoding with power re-scaling to satisfy the SINR thresholds. The ZF precoding matrix over a transmission block is given by
\begin{equation}
\tilde{\mathbf{W}}_{ZF}=\sqrt{\gamma}\sigma\tilde{\mathbf{H}}^H\left(\tilde{\mathbf{H}}\tilde{\mathbf{H}}^H\right)^{-1}.
\end{equation}
\item `BLP': Traditional block-level PM interference mitigation precoding in \cite{1561584}, which is solved by the fixed-point method.
\item `SV-CVX': Block-level PM slot-variant IE precoding, which is decomposed into $N_s$ PM-SLP subproblems based on Section \ref{secRela}. The subproblems are solved using CVX \cite{grant2014cvx}.
\item `SV-QP': The above subproblems are converted into their Lagrangian dual problems, then solved by the `quadprog' function in MATLAB.
\item `SI-CVX': Block-level PM slot-invariant IE precoding, which is solved using CVX \cite{grant2014cvx}.
\item `SI-QP': The above problem is converted into its Lagrangian dual problem, which is solved by the `quadprog' function in MATLAB.
\end{enumerate}

In simulations for SB problems:
\begin{enumerate}
\item `RZF': Linear RZF precoding with block-level power normalization to satisfy the block-level transmit power budget. The RZF precoding matrix over a transmission block is given by
\begin{equation}
\tilde{\mathbf{W}}_{RZF}=\frac{1}{f_{RZF}}\tilde{\mathbf{H}}^H\left(\tilde{\mathbf{H}}\tilde{\mathbf{H}}^H+\sigma^2\mathbf{I}\right)^{-1},
\end{equation}
where $f_{RZF}$ is a power scaling factor defined by
\begin{equation}
f_{RZF}=\frac{\left\|\tilde{\mathbf{H}}^H\left(\tilde{\mathbf{H}}\tilde{\mathbf{H}}^H+\sigma^2\mathbf{I}\right)^{-1}\tilde{\mathbf{S}}\right\|_{F}}{\sqrt{\sum^{N_s}_{i=1}p_i}}.
\end{equation}
\item `BLP': Traditional block-level SB interference mitigation precoding in \cite{1561584}, which is solved by the fixed-point method and the inverse property.
\item `SV-CVX': Block-level SB slot-variant IE precoding, which is decomposed into $N_s$ SB-SLP subproblems based on Section \ref{secRela}. The subproblems are solved using CVX \cite{grant2014cvx}. Theorem \ref{theorEquaSoluSB} is subsequently applied to compute the optimal solution to the original problem.
\item `SV-QP': The above SB-SLP subproblems are converted into PM-SLP subproblems, whose Lagrangian dual problems are solved by the `quadprog' function in MATLAB.
\item `SI-CVX': Block-level SB slot-invariant IE precoding, which is solved using CVX \cite{grant2014cvx}. 
\item `SI-QP': The block-level SB slot-invariant IE precoding problem is converted to its PM counterpart based on Section \ref{secDualSI}. The Lagrangian dual problem of the block-level PM slot-invariant IE precoding problem is solved by the `quadprog' function in MATLAB.
\end{enumerate}
	
\begin{figure}[!t]
\centering
\includegraphics[width=18pc]{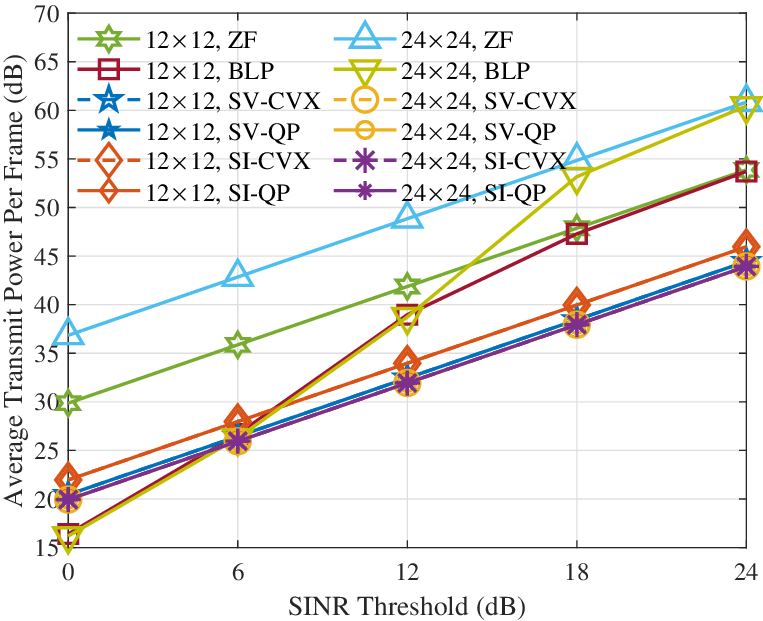}
\caption{Average transmit power per frame versus SINR threshold, $N_c=100$, $N_s=20$, QPSK.}
\label{fig_POW}
\end{figure}

Fig. \ref{fig_POW} compares the average block-level transmit power of various precoding schemes amongst different SINR thresholds for PM problems. The transmit power of the QP scheme is demonstrated to be consistent with its original problem solved by CVX, thus highlighting the effectiveness of the proposed optimal structure. It can be seen that, when the number of symbol slots is larger than the number of users, the PM slot-variant IE precoding always has the lowest block-level transmit power. The power gain comes from the joint optimization without the matrix structure constraint over a transmission block.

\begin{figure}[!t]
\centering
\includegraphics[width=18pc]{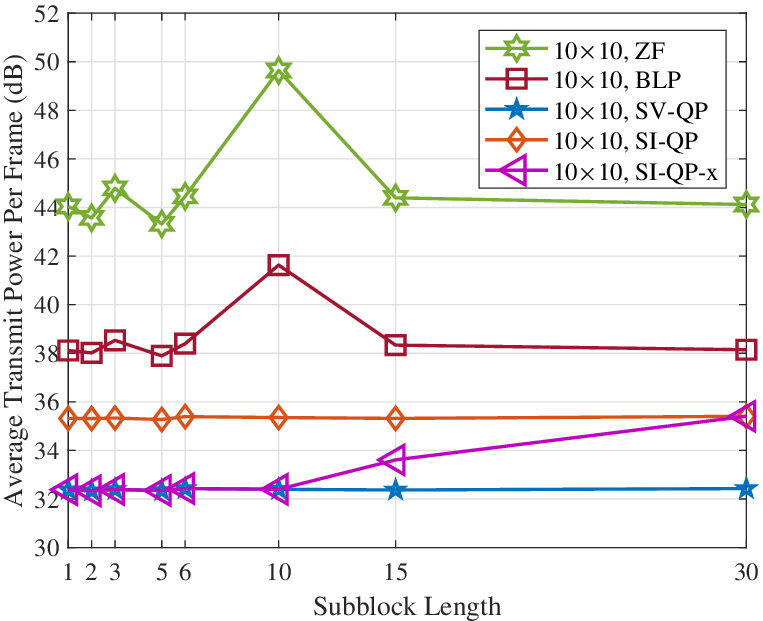}
\caption{Average transmit power per frame versus subblock length, $\gamma = 12 ~\text{dB}$, $Nc=2000$, $N_s=30$, QPSK.}
\label{fig_POWnsub}
\end{figure}

To further demonstrate the proposed precoders, we divided each transmission block/frame into several subblocks. For the PM problem, we denote `SI-QP-x' to represent that `SI-QP' adopted with a subblock of length `x'. Fig. \ref{fig_POWnsub} plots the average transmit power per frame as a function of the subblock length. It can be observed that when the subblock length does not exceed the number of users, `SI-QP-x' exhibits the same transmit power performance as `SV-QP'. However, when the subblock length exceeds the number of users, a larger subblock results in higher transmit power for `SI-QP-x'.

\begin{figure}[!t]
\centering
\includegraphics[width=18pc]{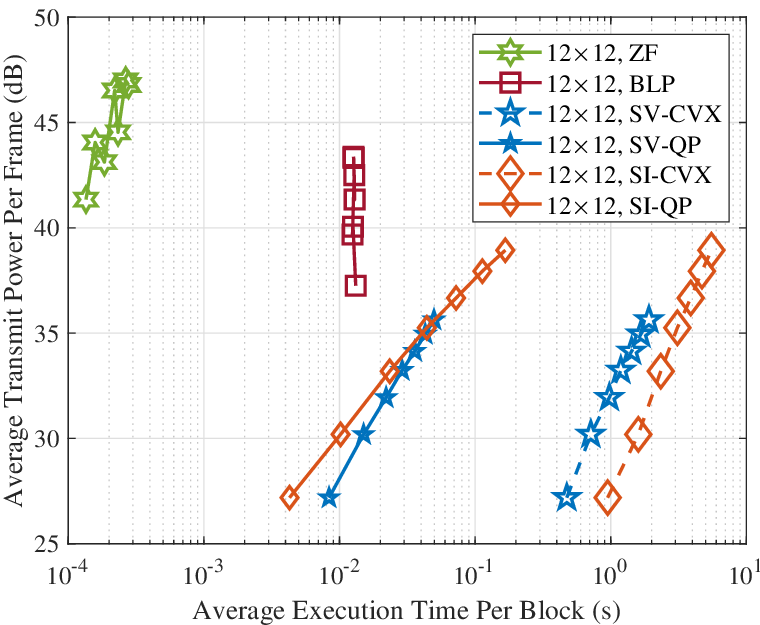}
\caption{Average transmit power versus average execution time per block, $N_c=2000$, $\gamma=12 ~\text{dB}$, $N_s=\left\{6, 12, 18, 24, 30, 36, 42\right\}$, QPSK.}
\label{fig_POWDurN}
\end{figure}

Fig. \ref{fig_POWDurN} demonstrates the performance-complexity tradeoff in terms of average transmit power and average execution time per block. The 7 data pairs for each scheme are obtained by evenly varying the block length from $N_s=6$ to $N_s=42$. We observe that the proposed block-level PM slot-variant and slot-invariant IE precoding schemes achieve lower transmit power compared to ZF and traditional BLP, although they generally require more execution time. When the block length is less than 12, the algorithms computed by the QP solver consume less execution time than traditional BLP.

\begin{figure}[!t]
\centering
\includegraphics[width=18pc]{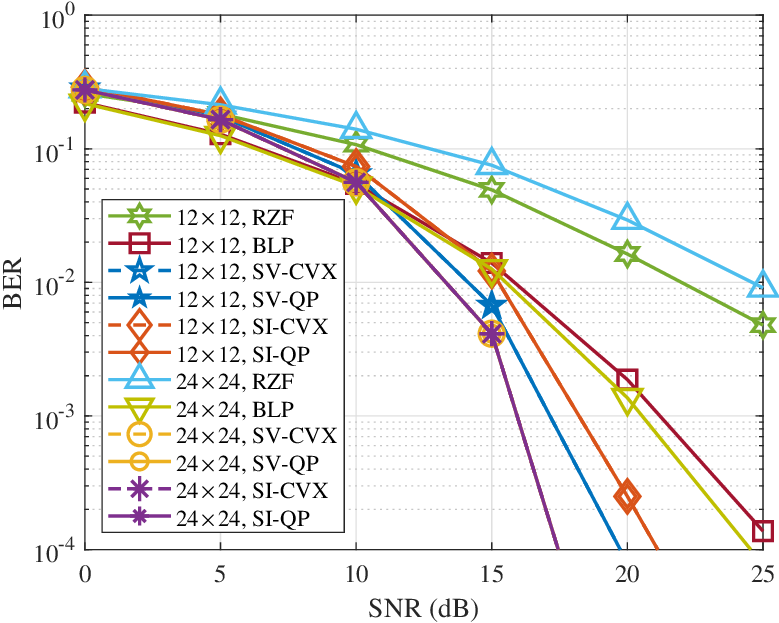}
\caption{BER versus SNR, $N_c=500$, $N_s=20$, QPSK.}
\label{fig_BER}
\end{figure}

Fig. \ref{fig_BER} compares the BER performance of the considered SB precoding schemes for different SNR. It shows that when the number of symbol slots exceeds the number of users, the proposed SB slot-variant IE precoding has the best BER performance due to its joint optimization without the matrix structure constraint over a transmission block/frame. 

\begin{figure}[!t]
\centering
\includegraphics[width=18pc]{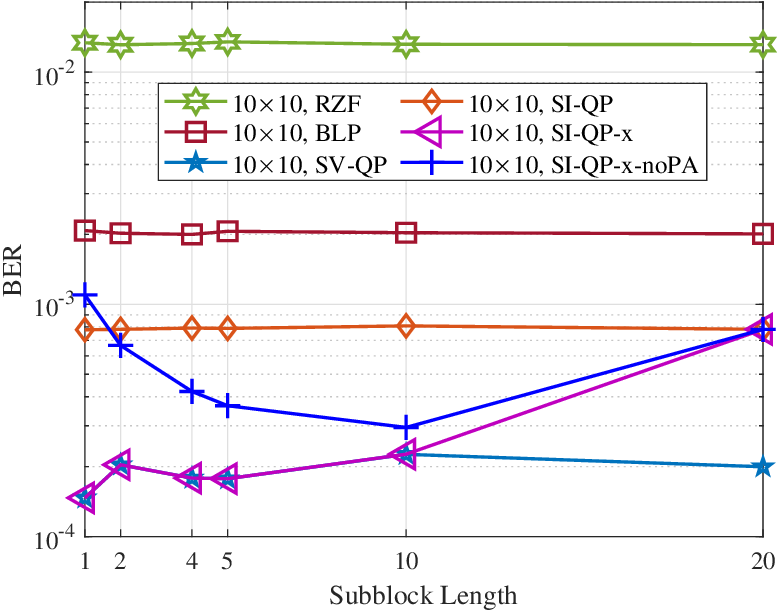}
\caption{BER versus subblock length, $N_c=10000$, $\text{SNR}=20 ~\text{dB}$, QPSK.}
\label{fig_BERnsub}
\end{figure}

Fig. \ref{fig_BERnsub} depicts the BER performance as a function of the subframe length. For each subframe in the `SI-QP-x' approach, we apply the SB slot-invariant IE precoding and the power allocation scheme described in Theorem \ref{theorEquaSoluSB}. In comparison, we also include the BER performance of the `SI-QP-x' method without power allocation, denoted as `SI-QP-x-noPA'. Fig. \ref{fig_BERnsub} demonstrates that when the subblock length does not exceed the number of users, `SI-QP-x' exhibits the same BER performance as `SV-QP' due to the presence of an intrinsic linear precoding structure for IE. As the subblock length increases, there is a tradeoff between BER performance and subblock length. Furthermore, this figure illustrates that power allocation within subblocks indeed enhances the BER performance. For `SI-QP-x-noPA', a valley point is observed  when the subblock length equals the number of users.

\begin{figure}[!t]
\centering
\includegraphics[width=18pc]{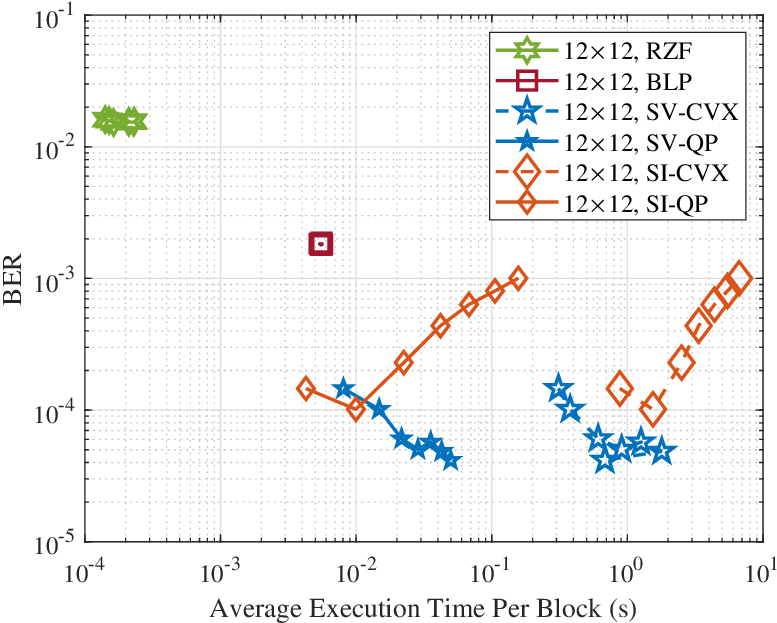}
\caption{BER versus average execution time per block, $N_c=20000$, $\text{SNR}=20 ~\text{dB}$, $N_s=\left\{6, 12, 18, 24, 30, 36, 42\right\}$, QPSK.}
\label{fig_BERDurN}
\end{figure}

Fig. \ref{fig_BERDurN} depicts the BER performance of the considered precoding algorithms in relation to the per block average execution time over a range of block lengths from $N_s=6$ to $N_s=42$, demonstrating a direct performance-complexity tradeoff. The block lengths are set to $\left\{6, 12, 18, 24, 30, 36, 42\right\}$. It is observed that the execution time of the proposed slot-invariant and slot-variant IE precoding schemes generally increases with block length, as it requires more calculations to solve optimization problems with larger dimensions. Furthermore, the results show that the QP solver, aided by the proposed optimal structure and explicit duality, can solve the problems more efficiently thane CVX.
	
\begin{figure}[!t]
\centering
\includegraphics[width=18pc]{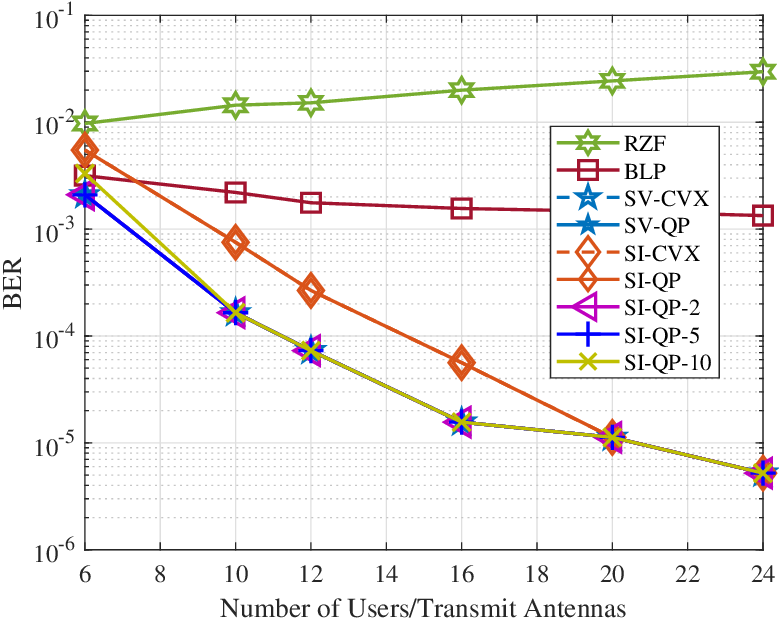}
\caption{BER versus number of users/transmit antennas, $N_c=1000$, $N_s=20$, $\text{SNR}=20 ~\text{dB}$, QPSK.}
\label{fig_BERk}
\end{figure}

Fig. \ref{fig_BERk} presents the BER performance of various SB precoding schemes as a function of the number of users in fully loaded MU-MISO systems. The figure shows that when the number of symbol slots in each transmission block exceeds the number of users, the slot-variant IE precoder demonstrates better BER performance compared to the slot-invariant IE precoder. The performance gap between them is almost proportional to the difference in the number of symbol slots and users. This performance gap disappears when the number of symbol slots equals or is less than the number of users. In scenarios where we divide a transmission block into several subframes, it is observed that when the subframe length does not exceed the number of users, `SI-QP-x' can achieve the same BER performance as the slot-variant IE precoding, thanks to the power allocation scheme.

\begin{figure}[!t]
\centering
\includegraphics[width=18pc]{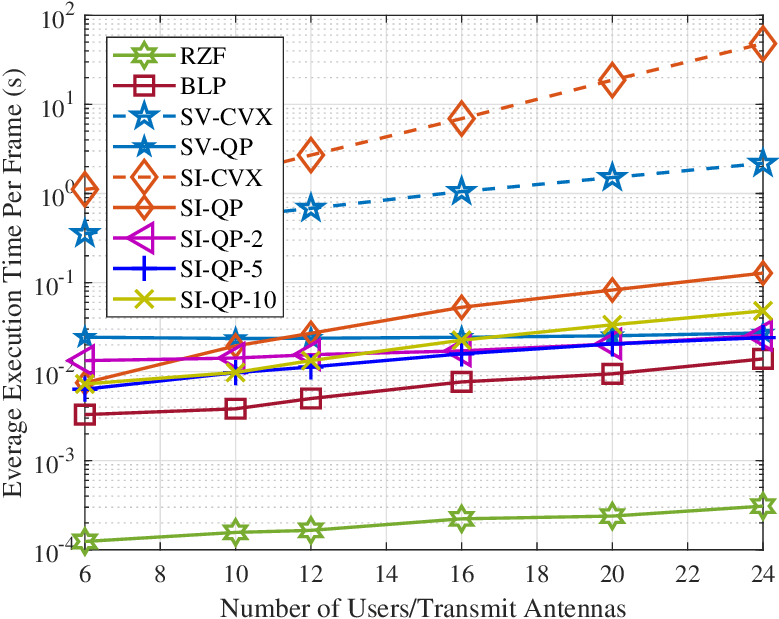}
\caption{Average execution time versus number of users/transmit antennas, $N_c=1000$, $N_s=20$, $\text{SNR}=20 ~\text{dB}$, QPSK.}
\label{fig_DURk}
\end{figure}

Fig. \ref{fig_DURk} illustrates the average execution time per frame of the SB precoding schemes presented in Fig. \ref{fig_BERk}. By employing the proposed optimal structure and explicit duality, the QP approach exhibits superior efficiency compared to the original problem solved by CVX, for both slot-invariant and slot-variant IE precoders. Furthermore, the slot-variant IE precoder requires far less processing time than the slot-invariant IE precoder because of its less-constrained formulation enabling us to decompose the problem into several subproblems with smaller variable sizes, which suggests that we prefer the slot-variant IE precoder in the scenario where the number of symbol slots does not exceed the number of users. This validates our idea of the linear precoder for IE in Section \ref{secLinear}. When dividing a transmission block into several subframes, with each subframe utilizing a constant precoder, the execution time of `SI-QP-x' is generally shorter than that of `SI-QP'. This observation can be interpreted as a performance-complexity tradeoff and indicates that by compromising the frame length, we can enhance the efficiency of the slot-invariant IE precoding scheme.

\section{Conclusion}
\label{secConclusion}
In this paper, we have presented block-level slot-variant and slot-invariant IE precoding. Both PM and weighted SB problems have been investigated, leveraging explicit duality. A linear precoder for IE has been further proposed, which can provide uncompromised SLP performance gain. Simulation results have been conducted to validate the effectiveness of the proposed slot-variant and slot-invariant IE precoding.Future works could involve devising practical numerical algorithms for the derived dual problems.

\appendices
\section{Proof for Proposition \ref{propDualFunc}}
\label{proofDualFunc}
For notational simplicity, define $\boldsymbol{\Sigma}\triangleq \sum_{i=1}^{N_s}\mathbf{s}_{:,i}\mathbf{s}^T_{:,i}+\mathbf{P}_3\mathbf{s}_{:,i}\mathbf{s}^T_{:,i}\mathbf{P}^T_3$. Substituting the optimal solution structure in \eqref{eqSoluStruc} into $\sum_{i=1}^{N_s}\mathbf{s}^T_{:,i}\overline{\mathbf{W}}^T\overline{\mathbf{W}}\mathbf{s}_{:,i}$ and $\sum_{i=1}^{N_s}\mathbf{s}^T_{:,i}\mathbf{P}^T_3\overline{\mathbf{W}}^T\overline{\mathbf{W}}\mathbf{P}_3\mathbf{s}_{:,i}$, we have the following two equations:
\begin{align}
\label{eqSWWS}
&\sum_{i=1}^{N_s}\mathbf{s}^T_{:,i}\overline{\mathbf{W}}^T\overline{\mathbf{W}}\mathbf{s}_{:,i}\nonumber\\
=&\frac{1}{4}\sum_{i=1}^{N_s}\sum_{j=1}^{N_s}\sum_{k=1}^{N_s}
\left(\mathbf{s}^T_{:,k}\frac{1}{\boldsymbol{\Sigma}}\mathbf{s}_{:,i}\mathbf{s}^T_{:,i}\frac{1}{\boldsymbol{\Sigma}}\mathbf{s}_{:,j}\right)\times \nonumber\\
& \boldsymbol{\lambda}^T_j\mathbf{T}\hat{\mathbf{S}}_j\mathbf{H}\mathbf{P}_1\mathbf{P}^T_1\mathbf{H}^T\hat{\mathbf{S}}^T_k\mathbf{T}^T\boldsymbol{\lambda}_k\nonumber\\
&+\frac{1}{4}\sum_{i=1}^{N_s}\sum_{j=1}^{N_s}\sum_{k=1}^{N_s}
\left(\mathbf{s}^T_{:,k}\mathbf{P}^T_3\frac{1}{\boldsymbol{\Sigma}}\mathbf{s}_{:,i}\mathbf{s}^T_{:,i}\frac{1}{\boldsymbol{\Sigma}}\mathbf{s}_{:,j}\right)\times \nonumber\\
& \boldsymbol{\lambda}^T_j\mathbf{T}\hat{\mathbf{S}}_j\mathbf{H}\mathbf{P}_1\mathbf{P}^T_2\mathbf{H}^T\hat{\mathbf{S}}^T_k\mathbf{T}^T\boldsymbol{\lambda}_k\nonumber\\
& +\frac{1}{4}\sum_{i=1}^{N_s}\sum_{j=1}^{N_s}\sum_{k=1}^{N_s}
\left(\mathbf{s}^T_{:,k}\frac{1}{\boldsymbol{\Sigma}}\mathbf{s}_{:,i}\mathbf{s}^T_{:,i}\frac{1}{\boldsymbol{\Sigma}}\mathbf{P}_3\mathbf{s}_{:,j}\right)\times \nonumber\\
& \boldsymbol{\lambda}^T_j\mathbf{T}\hat{\mathbf{S}}_j\mathbf{H}\mathbf{P}_2\mathbf{P}^T_1\mathbf{H}^T\hat{\mathbf{S}}^T_k\mathbf{T}^T\boldsymbol{\lambda}_k\nonumber\\
&+\frac{1}{4}\sum_{i=1}^{N_s}\sum_{j=1}^{N_s}\sum_{k=1}^{N_s}
\left(\mathbf{s}^T_{:,k}\mathbf{P}^T_3\frac{1}{\boldsymbol{\Sigma}}\mathbf{s}_{:,i}\mathbf{s}^T_{:,i}\frac{1}{\boldsymbol{\Sigma}}\mathbf{P}_3\mathbf{s}_{:,j}\right)\times \nonumber\\
& \boldsymbol{\lambda}^T_j\mathbf{T}\hat{\mathbf{S}}_j\mathbf{H}\mathbf{P}_2\mathbf{P}^T_2\mathbf{H}^T\hat{\mathbf{S}}^T_k\mathbf{T}^T\boldsymbol{\lambda}_k,
\end{align}
\begin{align}
\label{eqSPWWPS}
& \sum_{i=1}^{N_s}\mathbf{s}^T_{:,i}\mathbf{P}^T_3\overline{\mathbf{W}}^T\overline{\mathbf{W}}\mathbf{P}_3\mathbf{s}_{:,i}\nonumber\\
=&\frac{1}{4}\sum_{i=1}^{N_s}\sum_{j=1}^{N_s}\sum_{k=1}^{N_s}
\left(\mathbf{s}^T_{:,k}\frac{1}{\boldsymbol{\Sigma}}\mathbf{P}_3\mathbf{s}_{:,i}\mathbf{s}^T_{:,i}\mathbf{P}^T_3\frac{1}{\boldsymbol{\Sigma}}\mathbf{s}_{:,j}\right)\times \nonumber\\
& \boldsymbol{\lambda}^T_j\mathbf{T}\hat{\mathbf{S}}_j\mathbf{H}\mathbf{P}_1\mathbf{P}^T_1\mathbf{H}^T\hat{\mathbf{S}}^T_k\mathbf{T}^T\boldsymbol{\lambda}_k\nonumber\\
&+\frac{1}{4}\sum_{i=1}^{N_s}\sum_{j=1}^{N_s}\sum_{k=1}^{N_s}
\left(\mathbf{s}^T_{:,k}\mathbf{P}^T_3\frac{1}{\boldsymbol{\Sigma}}\mathbf{P}_3\mathbf{s}_{:,i}\mathbf{s}^T_{:,i}\mathbf{P}^T_3\frac{1}{\boldsymbol{\Sigma}}\mathbf{s}_{:,j}\right)\times \nonumber\\
& \boldsymbol{\lambda}^T_j\mathbf{T}\hat{\mathbf{S}}_j\mathbf{H}\mathbf{P}_1\mathbf{P}^T_2\mathbf{H}^T\hat{\mathbf{S}}^T_k\mathbf{T}^T\boldsymbol{\lambda}_k\nonumber\\
&+\frac{1}{4}\sum_{i=1}^{N_s}\sum_{j=1}^{N_s}\sum_{k=1}^{N_s}
\left(\mathbf{s}^T_{:,k}\frac{1}{\boldsymbol{\Sigma}}\mathbf{P}_3\mathbf{s}_{:,i}\mathbf{s}^T_{:,i}\mathbf{P}^T_3\frac{1}{\boldsymbol{\Sigma}}\mathbf{P}_3\mathbf{s}_{:,j}\right)\times \nonumber\\
& \boldsymbol{\lambda}^T_j\mathbf{T}\hat{\mathbf{S}}_j\mathbf{H}\mathbf{P}_2\mathbf{P}^T_1\mathbf{H}^T\hat{\mathbf{S}}^T_k\mathbf{T}^T\boldsymbol{\lambda}_k\nonumber\\
&+\frac{1}{4}\sum_{i=1}^{N_s}\sum_{j=1}^{N_s}\sum_{k=1}^{N_s}
\left(\mathbf{s}^T_{:,k}\mathbf{P}^T_3\frac{1}{\boldsymbol{\Sigma}}\mathbf{P}_3\mathbf{s}_{:,i}\mathbf{s}^T_{:,i}\mathbf{P}^T_3\frac{1}{\boldsymbol{\Sigma}}\mathbf{P}_3\mathbf{s}_{:,j}\right)\times \nonumber\\
& \boldsymbol{\lambda}^T_j\mathbf{T}\hat{\mathbf{S}}_j\mathbf{H}\mathbf{P}_2\mathbf{P}^T_2\mathbf{H}^T\hat{\mathbf{S}}^T_k\mathbf{T}^T\boldsymbol{\lambda}_k.
\end{align}
To simplify the above expressions, we can rearrange components and utilize the definition of $\boldsymbol{\Sigma}$, resulting in
\begin{align}
\label{eqSPWWPS2}
&\sum_{i=1}^{N_s}\mathbf{s}^T_{:,i}\overline{\mathbf{W}}^T\overline{\mathbf{W}}\mathbf{s}_{:,i}+\sum_{i=1}^{N_s}\mathbf{s}^T_{:,i}\mathbf{P}^T_3\overline{\mathbf{W}}^T\overline{\mathbf{W}}\mathbf{P}_3\mathbf{s}_{:,i}\nonumber\\
=&\frac{1}{4}\sum_{j=1}^{N_s}\sum_{k=1}^{N_s}
\left(\mathbf{s}^T_{:,k}\frac{1}{\boldsymbol{\Sigma}}\mathbf{s}_{:,j}\right)
\boldsymbol{\lambda}^T_j\mathbf{T}\hat{\mathbf{S}}_j\mathbf{H}\mathbf{P}_1\mathbf{P}^T_1\mathbf{H}^T\hat{\mathbf{S}}^T_k\mathbf{T}^T\boldsymbol{\lambda}_k\nonumber\\
&+\frac{1}{4}\sum_{j=1}^{N_s}\sum_{k=1}^{N_s}
\left(\mathbf{s}^T_{:,k}\mathbf{P}^T_3\frac{1}{\boldsymbol{\Sigma}}\mathbf{s}_{:,j}\right)
\boldsymbol{\lambda}^T_j\mathbf{T}\hat{\mathbf{S}}_j\mathbf{H}\mathbf{P}_1\mathbf{P}^T_2\mathbf{H}^T\hat{\mathbf{S}}^T_k\mathbf{T}^T\boldsymbol{\lambda}_k\nonumber\\
&+\frac{1}{4}\sum_{j=1}^{N_s}\sum_{k=1}^{N_s}
\left(\mathbf{s}^T_{:,k}\frac{1}{\boldsymbol{\Sigma}}\mathbf{P}_3\mathbf{s}_{:,j}\right)
\boldsymbol{\lambda}^T_j\mathbf{T}\hat{\mathbf{S}}_j\mathbf{H}\mathbf{P}_2\mathbf{P}^T_1\mathbf{H}^T\hat{\mathbf{S}}^T_k\mathbf{T}^T\boldsymbol{\lambda}_k\nonumber\\
&+\frac{1}{4}\sum_{j=1}^{N_s}\sum_{k=1}^{N_s}
\left(\mathbf{s}^T_{:,k}\mathbf{P}^T_3\frac{1}{\boldsymbol{\Sigma}}\mathbf{P}_3\mathbf{s}_{:,j}\right)\times\nonumber\\
&\boldsymbol{\lambda}^T_j\mathbf{T}\hat{\mathbf{S}}_j\mathbf{H}\mathbf{P}_2\mathbf{P}^T_2\mathbf{H}^T\hat{\mathbf{S}}^T_k\mathbf{T}^T\boldsymbol{\lambda}_k.
\end{align}
Subsequently, we use the optimal solution structure to rewrite the remaining terms of the Lagrangian, yielding
\begin{align}
\label{eqSPWWPS3}
&\sum_{i=1}^{N_s}\boldsymbol{\lambda}^T_i\mathbf{T}\hat{\mathbf{S}}_i\mathbf{H}\mathbf{P}_1\overline{\mathbf{W}}\mathbf{s}_{:,i}+\boldsymbol{\lambda}^T_i\mathbf{T}\hat{\mathbf{S}}_i\mathbf{H}\mathbf{P}_2\overline{\mathbf{W}}\mathbf{P}_3\mathbf{s}_{:,i}\nonumber\\
=&\frac{1}{2}\sum_{i=1}^{N_s}\sum_{j=1}^{N_s}
\left(\mathbf{s}^T_{:,j}\frac{1}{\boldsymbol{\Sigma}}\mathbf{s}_{:,i}\right)
\boldsymbol{\lambda}^T_i\mathbf{T}\hat{\mathbf{S}}_i\mathbf{H}\mathbf{P}_1\mathbf{P}^T_1\mathbf{H}^T\hat{\mathbf{S}}^T_j\mathbf{T}^T\boldsymbol{\lambda}_j\nonumber\\
&+\frac{1}{2}\sum_{i=1}^{N_s}\sum_{j=1}^{N_s}
\left(\mathbf{s}^T_{:,j}\mathbf{P}^T_3\frac{1}{\boldsymbol{\Sigma}}\mathbf{s}_{:,i}\right)
\boldsymbol{\lambda}^T_i\mathbf{T}\hat{\mathbf{S}}_i\mathbf{H}\mathbf{P}_1\mathbf{P}^T_2\mathbf{H}^T\hat{\mathbf{S}}^T_j\mathbf{T}^T\boldsymbol{\lambda}_j\nonumber\\
&+\frac{1}{2}\sum_{i=1}^{N_s}\sum_{j=1}^{N_s}
\left(\mathbf{s}^T_{:,j}\frac{1}{\boldsymbol{\Sigma}}\mathbf{P}_3\mathbf{s}_{:,i}\right)
\boldsymbol{\lambda}^T_i\mathbf{T}\hat{\mathbf{S}}_i\mathbf{H}\mathbf{P}_2\mathbf{P}^T_1\mathbf{H}^T\hat{\mathbf{S}}^T_j\mathbf{T}^T\boldsymbol{\lambda}_j\nonumber\\
&+\frac{1}{2}\sum_{i=1}^{N_s}\sum_{j=1}^{N_s}
\left(\mathbf{s}^T_{:,j}\mathbf{P}^T_3\frac{1}{\boldsymbol{\Sigma}}\mathbf{P}_3\mathbf{s}_{:,i}\right)\times\nonumber\\
&\boldsymbol{\lambda}^T_i\mathbf{T}\hat{\mathbf{S}}_i\mathbf{H}\mathbf{P}_2\mathbf{P}^T_2\mathbf{H}^T\hat{\mathbf{S}}^T_j\mathbf{T}^T\boldsymbol{\lambda}_j.
\end{align}
After the above manipulations, the dual function is given by
\begin{align*}
& g\left(\left\{\boldsymbol{\lambda}_i\right\}\right)\nonumber\\
\triangleq&\min_{\overline{\mathbf{W}}}\mathcal{L}\left(\overline{\mathbf{W}},\right\{\boldsymbol{\lambda}_i\left\}\right)\nonumber\\
=&-\frac{1}{4}\sum_{i=1}^{N_s}\sum_{j=1}^{N_s}
\left(\mathbf{s}^T_{:,j}\frac{1}{\boldsymbol{\Sigma}}\mathbf{s}_{:,i}\right)
\boldsymbol{\lambda}^T_i\mathbf{T}\hat{\mathbf{S}}_i\mathbf{H}\mathbf{P}_1\mathbf{P}^T_1\mathbf{H}^T\hat{\mathbf{S}}^T_j\mathbf{T}^T\boldsymbol{\lambda}_j\nonumber\\
&-\frac{1}{4}\sum_{i=1}^{N_s}\sum_{j=1}^{N_s}
\left(\mathbf{s}^T_{:,j}\mathbf{P}^T_3\frac{1}{\boldsymbol{\Sigma}}\mathbf{s}_{:,i}\right)
\boldsymbol{\lambda}^T_i\mathbf{T}\hat{\mathbf{S}}_i\mathbf{H}\mathbf{P}_1\mathbf{P}^T_2\mathbf{H}^T\hat{\mathbf{S}}^T_j\mathbf{T}^T\boldsymbol{\lambda}_j\nonumber\\
&-\frac{1}{4}\sum_{i=1}^{N_s}\sum_{j=1}^{N_s}
\left(\mathbf{s}^T_{:,j}\frac{1}{\boldsymbol{\Sigma}}\mathbf{P}_3\mathbf{s}_{:,i}\right)
\boldsymbol{\lambda}^T_i\mathbf{T}\hat{\mathbf{S}}_i\mathbf{H}\mathbf{P}_2\mathbf{P}^T_1\mathbf{H}^T\hat{\mathbf{S}}^T_j\mathbf{T}^T\boldsymbol{\lambda}_j\nonumber\\
\end{align*}
\begin{align}
\label{eqDualFuncDerive}
&-\frac{1}{4}\sum_{i=1}^{N_s}\sum_{j=1}^{N_s}
\left(\mathbf{s}^T_{:,j}\mathbf{P}^T_3\frac{1}{\boldsymbol{\Sigma}}\mathbf{P}_3\mathbf{s}_{:,i}\right)
\times\nonumber\\
& \boldsymbol{\lambda}^T_i\mathbf{T}\hat{\mathbf{S}}_i\mathbf{H}\mathbf{P}_2\mathbf{P}^T_2\mathbf{H}^T\hat{\mathbf{S}}^T_j\mathbf{T}^T\boldsymbol{\lambda}_j+\sum_{i=1}^{N_s}\boldsymbol{\lambda}^T_i\mathbf{b}_i\nonumber\\
=&-\frac{1}{4}\boldsymbol{\lambda}^T\mathbf{U}\boldsymbol{\lambda}+\boldsymbol{\lambda}^T\mathbf{b},
\end{align}
where
\begin{subequations}
\begin{align}
\boldsymbol{\lambda}\triangleq & \left[\boldsymbol{\lambda}^T_1, \cdots, \boldsymbol{\lambda}^T_{N_s}\right]^T\in \mathbb{R}^{2N_s N_r},\\
\mathbf{U}\triangleq & \begin{bmatrix}
\mathbf{U}_{1,1}&\cdots &\mathbf{U}_{1,N_s}\\
\vdots & \ddots & \vdots\\
\mathbf{U}_{N_s,1} & \cdots & \mathbf{U}_{N_s,N_s}
\end{bmatrix}\in \mathbb{R}^{2N_s N_r \times 2N_s N_r},\\
\mathbf{U}_{i,j}\triangleq &
\left(\mathbf{s}^T_{:,j}\frac{1}{\boldsymbol{\Sigma}}\mathbf{s}_{:,i}\right)
\mathbf{T}\hat{\mathbf{S}}_i\mathbf{H}\mathbf{P}_1\mathbf{P}^T_1\mathbf{H}^T\hat{\mathbf{S}}^T_j\mathbf{T}^T\nonumber\\
&+\left(\mathbf{s}^T_{:,j}\mathbf{P}^T_3\frac{1}{\boldsymbol{\Sigma}}\mathbf{s}_{:,i}\right)
\mathbf{T}\hat{\mathbf{S}}_i\mathbf{H}\mathbf{P}_1\mathbf{P}^T_2\mathbf{H}^T\hat{\mathbf{S}}^T_j\mathbf{T}^T\nonumber\\
&+\left(\mathbf{s}^T_{:,j}\frac{1}{\boldsymbol{\Sigma}}\mathbf{P}_3\mathbf{s}_{:,i}\right)
\mathbf{T}\hat{\mathbf{S}}_i\mathbf{H}\mathbf{P}_2\mathbf{P}^T_1\mathbf{H}^T\hat{\mathbf{S}}^T_j\mathbf{T}^T\nonumber\\
&+\left(\mathbf{s}^T_{:,j}\mathbf{P}^T_3\frac{1}{\boldsymbol{\Sigma}}\mathbf{P}_3\mathbf{s}_{:,i}\right)
\mathbf{T}\hat{\mathbf{S}}_i\mathbf{H}\mathbf{P}_2\mathbf{P}^T_2\mathbf{H}^T\hat{\mathbf{S}}^T_j\mathbf{T}^T.
\end{align}
\end{subequations}

\bibliographystyle{IEEEtran}
\bibliography{IEEEabrv,references}

\end{document}